\documentclass[aps, prb, twocolumn, superscriptaddress, floatfix]{revtex4-2}
\usepackage{graphicx}
\usepackage{amsmath}
\usepackage{amssymb}
\usepackage{tikz}
\usepackage{braket}
\usepackage{pgfplots}
\usepackage[version=4]{mhchem}
\pgfplotsset{compat=1.17}
\usepackage{subcaption}
\usetikzlibrary{shapes.geometric}
\usetikzlibrary{shadows,patterns,perspective}
\usetikzlibrary {decorations,decorations.text}
\usetikzlibrary{decorations.pathreplacing}
\usepackage{float}
\usepackage{hyperref}
\usepackage{xcolor,ulem}
        
\definecolor{linkcolor}{RGB}{0, 0, 255}      
\definecolor{citecolor}{RGB}{0, 128, 0}     
\definecolor{urlcolor}{RGB}{255, 0, 0}       

\hypersetup{
    colorlinks=true,
    linkcolor=linkcolor,
    citecolor=citecolor,
    urlcolor=urlcolor,
    linktoc=all,  
    pdfborder={0 0 0}  
}

\DeclareCaptionJustification{justified}{\leftskip=0pt \rightskip=0pt \parfillskip=0pt plus 1fil}

\begin{document}
\definecolor{dy}{rgb}{0.9,0.9,0.4}
\definecolor{dr}{rgb}{0.95,0.65,0.55}
\definecolor{db}{rgb}{0.5,0.8,0.9}
\definecolor{dg}{rgb}{0.2,0.9,0.6}
\definecolor{BrickRed}{rgb}{0.8,0.3,0.3}
\definecolor{Navy}{rgb}{0.2,0.2,0.6}
\definecolor{DarkGreen}{rgb}{0.1,0.4,0.1}

\title{Quantum Zeno Effect in Noisy Integrable Quantum Circuits for Impurity Models}

\author{Yicheng Tang}
\email{tang.yicheng@rutgers.edu}
\affiliation{Department of Physics and Astronomy, Center for Materials Theory, Rutgers University,
Piscataway, New Jersey 08854, United States of America} 

\author{Pradip Kattel}
\affiliation{Department of Physics and Astronomy, Center for Materials Theory, Rutgers University,
Piscataway, New Jersey 08854, United States of America} 

\author{J.~H.~Pixley}
\affiliation{Department of Physics and Astronomy, Center for Materials Theory, Rutgers University,
Piscataway, New Jersey 08854, United States of America} 
\affiliation{Center for Computational Quantum Physics, Flatiron Institute, 162 5th Avenue, New York, NY 10010}

\author{Natan Andrei}
\affiliation{Department of Physics and Astronomy, Center for Materials Theory, Rutgers University,
Piscataway, New Jersey 08854, United States of America}

\begin{abstract}
We theoretically study the open quantum system dynamics (in the Trotterized limit) of integrable quantum circuits in the presence of onsite dephasing noise with a spin-1/2 ``impurity'' interacting at the edge.  Using a combination of Bethe Ansatz (BA) and exact diagonalization (ED), we study the dynamics of both the bulk and the impurity for the XXX (Heisenberg) and the XX qubit chains in the presence and absence of bulk noise. In the absence of noise, we show that the impurity exhibits two distinct phases: the bound mode phase, where the impurity keeps oscillating in time, and the Kondo phase, where it decays with Kondo time $t_K$. Turning on the bulk dephasing noise, we find for the two models that in the long-time limit in both regimes, the quantum Zeno effect takes place where the dynamics of the impurity magnetization slows down as the noise strength $\gamma$ increases. The impurity magnetization in the bound-mode regime shows the opposite effect, decaying faster as the noise strength increases for short times ($t \ll 1/\gamma$). We show that the Zeno effect slowing down the impurity dynamics in the long-time limit is driven by the change in bulk dynamics from ballistic (KPZ) universality class in the XX (XXX) chain to diffusive dynamics in the presence of noise.



\end{abstract}

\maketitle

\textit{Introduction}: 
The interplay between noise and quantum coherence is central to understanding modern quantum devices, particularly in noisy intermediate-scale quantum (NISQ) platforms such as superconducting qubits~\cite{SCQkrantz2019quantum,SCQdevoret2004superconducting,SCQkjaergaard2020superconducting}, trapped ions~\cite{TIblatt2012quantum,TIbruzewicz2019trapped,TIleibfried2003quantum,TIsinger2010colloquium}, neutral atoms~\cite{NAhenriet2020quantum,NAjessen2004quantum,NAwintersperger2023neutral}, and circuit quantum electrodynamics~\cite{cQEDblais2020quantum,cQEDclerk2020hybrid,cQEDblais2021circuit,cQEDharoche2020cavity,cQEDschmidt2013circuit}. In these open systems~\cite{breuer2002theory,banerjee2018open,schaller2014open,mirrahimi2015dynamics}, noise—ranging from dephasing and relaxation to control errors—fundamentally limits coherence times and gate fidelities~\cite{porter2022impact,otten2021impacts}, making it imperative to study its effects on quantum many-body systems (QMBS). Classically simulating large QMBS is infeasible due to exponential Hilbert space growth~\cite{manin1980computable,feynman2018simulating,lloyd1996universal}, but quantum computers naturally accommodate such complexity~\cite{feynman2018simulating,lloyd1996universal}. A promising approach involves Trotterizing time evolution into quantum circuits~\cite{williams2010explorations,nielsen2010quantum}, with integrable circuits~\cite{ljubotina2019ballistic,paletta2024integrability,hudomal2024integrability,hutsalyuk2024exact,miao2023integrable,gombor2024integrable} offering unique advantages. Built from Yang-Baxter-solvable $R$-matrices~\cite{perk2006yang,jimbo1994introduction}, these circuits simulate integrable Hamiltonians~\cite{baxter2007exactly,reshetikhin2010lectures,jimbo1994algebraic,korepin1997quantum} with exact control over dynamics, enabling insights into complex many-body behavior~\cite{yu2025quantum,IQChutsalyuk2024exact,IQCmiao2023integrable,zhang2024observation}.

However, real-world quantum devices are inherently noisy~\cite{monroe2021programmable,QCaspuru2012photonic,QCbernien2017probing,QCblatt2012quantum,QCbloch2012quantum,QCgross2017quantum,QChouck2012chip,QCjoshi2022probing,QCsemeghini2021probing,QCzhang2017observation}, which disrupts transport and induces phenomena like dynamical phase transitions~\cite{kawabata2023dynamical,PhysRevB.108.075110}, the quantum Zeno effect (where increasing the strength of noise slows down quantum dynamic)~\cite{lerner2018quantum,vanhoecke2024kondo,javed2024zeno}, and the quantum Mpemba effect (where non-equilibrium quantum systems relax faster when they are further from equilibrium)~\cite{wang2024mpemba,chatterjee2023quantum,rylands2024microscopic,PhysRevLett.133.010402}. To disentangle these effects, we study an exactly solvable integrable circuit with dephasing noise, bridging the gap between ideal unitary dynamics and realistic noisy environments. Our work reveals how noise inherently modifies coherent evolution, offering a controlled framework to benchmark NISQ experiments~\cite{schlimgen2022quantum} by studying anisotropic Heisenberg spin chains which can be simulated on experimentally available platforms~\cite{van2021quantum,murmann2015antiferromagnetic}.



\begin{figure}
    \centering    \includegraphics[width=0.9\linewidth]{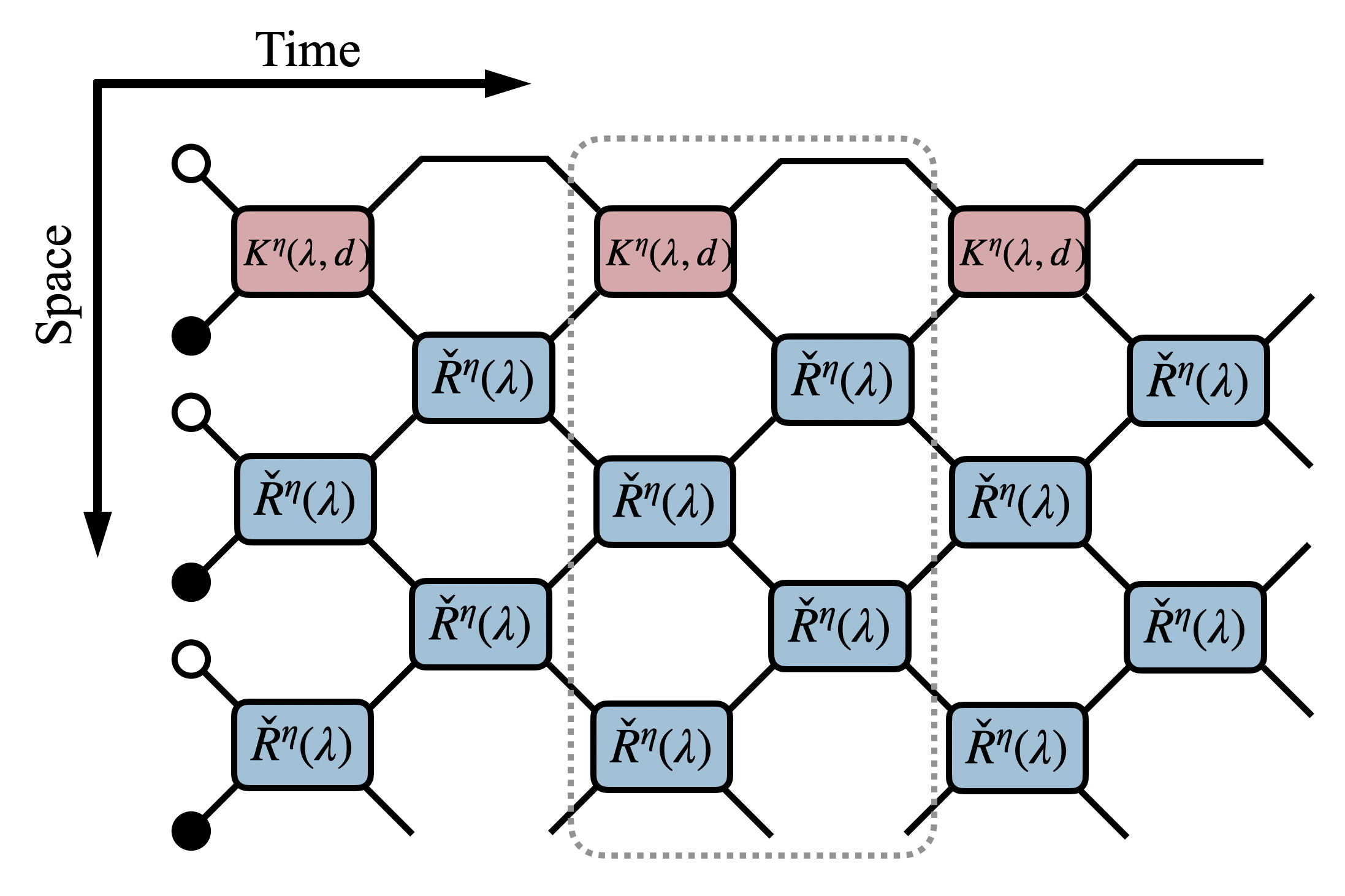}
    \begin{tikzpicture}[overlay, remember picture]
         \node[black] at (-3.6,4.7) {$\mathbb U^\eta(\lambda, d)$};
    \end{tikzpicture}
    \caption{Illustrations of an integrable quantum circuit with spin-1/2 qubits in an initial classical N\'eel state (empty circles are qubits in the up states and black filled circles are in the down qubit state). Each rectangle represents a local unitary matrix; blue ones are the braid form of the trigonometric six-vertex R-matrix $\check R(\lambda)$ and red ones are the boundary reflection matrix $K(\lambda,d) = \check R(-d)\check R(\lambda+d)$.
    The dashed rectangle shows one layer of the time evolution operator $\mathbb U^\eta(\lambda,d)$.  
    }
    \label{IntegrableCircut}
\end{figure}

\textit{Models:} To begin with, we construct an integrable quantum circuit as shown in Fig.\ref{IntegrableCircut} 
that corresponds to a unitary operator

\begin{equation}
\label{theUnitary}
\begin{split}
    \mathbb U^\eta (\lambda,d)
    &= K_{0,1}^\eta(\lambda,d)\prod_{j=1}^{L/2-1} \check R _{2j,2j+1}^\eta(\lambda) \check R_{2j-1,2j}^\eta(\lambda),
\end{split}
\end{equation}
where $L$ is the number of qubits, $\check{R}_{j,j+1}$ given explicitly in Appendix~\ref{mappingQC-Liouvillian} is the braid form of the trigonometric $R$-matrix of the six-vertex model~\cite{sutherland2004beautiful,baxter2007exactly} acting on the $j$ and $j+1$ qubits, $ K_{01}^\eta(\lambda,d)=  \check R_{01}^\eta(-d) \check R^\eta_{01}(\lambda+d)$ is the boundary $K$-matrix that solves the reflection algebra~\cite{cherednik1984factorizing,sklyanin1988boundary,wang2015off}, and $d$, $\lambda$ and $\eta$ are free parameters.
In the Trotterziation limit~\cite{trotter1959product,suzuki1991general},  choosing $\lambda = -i\delta t\sinh\eta$, this quantum circuit simulates the unitary evolution $\exp(-iHt)  = \lim_{\delta t \to 0}\mathbb U(\lambda,d,\eta)^{\frac{t}{\delta t}}  $~ for an integrable Hamiltonian $H$~\cite{lloyd1996universal,nielsen2010quantum,whitfield2011simulation},  where 
\begin{equation}
\begin{split}
   H=&\sum_{j=1}^{L-1}\frac{1}{2}(\sigma_j^x \sigma_{j+1}^x+\sigma_j^y \sigma_{j+1}^y+\Delta\sigma_j^z \sigma_{j+1}^z)\\
    &+\frac{J}{2} \left(\sigma_0^x \sigma_{1}^x+\sigma_0^y \sigma_{1}^y+\Delta'\sigma_0^z \sigma_{1}^z\right)\\
\end{split}
\label{ham}
\end{equation}
with $\Delta \mathord{=} \cosh(\eta)$, $J \mathord{=} \frac{\sinh^2(\eta)\cosh(d)}{\sinh^2(\eta)-\sinh^2(d)}$ and $\Delta'\mathord{=} \frac{\cosh(\eta)}{\cosh(d)} $. This Hamiltonian describes the $XXZ$ spin chain with a boundary impurity.
At the isotropic XXX ($\Delta\mathord{=}\Delta'\mathord{=}1)$ and the XX ($\Delta\mathord{=}\Delta'\mathord{=}0$) points, the model is integrable for arbitrary boundary coupling $J$. For generic $\Delta$, the boundary coupling ($J$) and the boundary anisotropy parameter ($\Delta'$) are related and expressed in terms of a single free parameter $d$ to maintain the integrability~\cite{XXZKondo,kattel2024kondo, wang1997exact, furusaki1998kondo,hu1998two,laflorencie2008kondo,kattel2023kondo}. In the ground state, the impurity is screened due to the Kondo effect when $\bar d\mathord{=}d/\eta $ is purely imaginary or $0<\bar d<\frac{1}{2}$ and it is screened by a bound mode exponentially localized at the boundary when $1>\bar d>\frac{1}{2}$ \cite{XXZKondo}.


To study the effect of dephasing noise on the integrable quantum circuit, we solve the continuous-time dynamics of the noisy integrable Hamiltonian using the Gorini-Kossakowski-Sudarshan-Lindblad (GKSL) equation, $\mathcal{L}[\rho(t)] = i\partial_t\rho(t)$~\cite{banerjee2018open,breuer2002theory} that governs the dynamics of the density matrix $\rho(t)$ with a Liouvillian~\cite{lindblad1976generators,gorini1976completely,ILPhysRevB.103.115132,ILsu2022integrable}
\begin{equation}
    \mathcal{L}(\rho)= -\left[\rho,H\right] 
+i \sum_{j=1}^{L-1}\gamma\left(L_j \rho L_j^\dagger-\frac{1}{2}\left\{L^\dagger_j L_j,\rho \right\} \right),
\end{equation}
where the jump operators are $L_j\mathord{=}\sigma^z_j$, $\gamma$ is the dephasing rate, $H$ is the Hamiltonian in Eq.~\eqref{ham}, $\{A,B\}$ and $[A,B]$
denote the anticommutator and the commutator, respectively.  In the following, we will evolve the system from two initial states, N\'eel state and domain wall state, expressed as: 
$\rho_{N/D} = (\prod_{z\in \mathcal{S}_{N/D}}\sigma^+_z)\ket{0}\bra{0}(\prod_{z\in \mathcal{S}_{N/D}}\sigma^-_z)$ 
where $\mathcal{S}_{N/D}$ denote the set of sites with $z \text{ }\mathord{=}\text{ } 0,2,4,...,L$ for the N\'eel (N) state $\rho_N$, and $z\text{ }\mathord{=}\text{ }0,1,2,...,L/2$ for the domain wall (D) state $\rho_D$.
Here, $\ket{0}$ refers to the state with all spins down (qubits $0$) state. In both cases, we compute the time-dependent expectation value of magnetization at site $j$, $S^z_j(t)\mathord{=}\mathrm{Tr}\{\sigma^z_j\rho(t)\}$. 

\textit{XX chain}: Let us start from the XX point as it allows for a simpler treatment with similar phenomenology~\cite{kattel2024kondo}.  It is convenient to work in the fermionic language via a Jordan-Wigner transformation  $c^\dagger_j\mathord{=}\sigma^+_j\prod_{l<j}\sigma^z_l $. The magnetization is then related to the two-point correlator $G(x,y;t) = \mathrm{Tr}[c^\dagger_{x}c_{y}\rho(t)]$ as $S^z_j(t)=2G(j,j;t)-1$. 
A closed set of equations gives the dynamics of the two-point correlators because the Hamiltonian and the jump operators are both quadratic in fermions~\cite{vznidarivc2010exact,Essler-prozen}. Therefore, the dynamics of the two-point correlation function in the noisy XX model can be mapped to a {two-particle} quantum mechanical problem as shown in Appendix~\ref{hubbardmap}. The correlator can be vectorized as a state (see Appendix~\ref{hubbardmap}) $|G(t))\in\mathbb{C}^{L^2}$ where  $|G(t))=\sum_{x,y}G(x,y;t)|x,y)$ and $|x,y)$ forms an orthonormal basis in $\mathbb{C}^{L^2}$. Its evolution from an initial state $|G_{in})=\sum_{x,y}\mathrm{Tr}[G(x,y{;0})\rho_{N/D}]|x,y)$ reads $|G(t))=e^{-i\hat{h}t}|G_{in})$ with Liouvillian $\hat{h}=\sum_{x,y,x',y'}h^{x,y}_{x',y'}|x,y)(x',y'|$ and
\begin{equation}
\begin{aligned}
 h^{x,y}_{x',y'}  &= J_{x}(\delta_{x,x'+1}+\delta_{x,x'-1})-J_{y}(\delta_{y,y'+1}+\delta_{y,y'-1})\\
&+4i\gamma(\delta_{x,x'}\delta_{y,y'}-1)+2i\gamma(\delta_{x',0}+\delta_{y',0}),
\label{fqham}
\end{aligned}
\end{equation}
where $J_x= 1+\delta_{x,0}(J-1)$.
The Liouvillian {right} eigenmodes
$|G_{E})=\sum_{x,y}G_E(x,y)|x,y)$ satisfy the stationary Sch\"odinger equation $\hat{h}|G_{E})=E|G_E)$ with eigenvalue $E$ in terms of which we expand the time-dependent state $|G(t)) = \sum_{E} (\bar{G}_E|G_{in})e^{-iEt}|G_E)$. Here $(\bar{G}_E|$ is the left eigenmode of the Liouvillian satisfying $(\bar{G}_E|\hat{h}^\dagger = E^*(\bar{G}_E|$ and it forms a bi-orthognal basis with $|G_E)$ as $\sum_E |G_E)(\bar{G}_E| = \mathbb{I}$.
The time-dependent impurity  magnetization (at site $j=0$) can be written as 
\begin{equation}
\label{TIM}
    S^z_0 (t)= 2\sum_{E}\alpha_{E}^{N/D}e^{-iEt}-1,
\end{equation}
where the amplitude $\alpha_{E}^{N/D}= G_E(0,0)(\bar{G}_E|G_{in})$ depends on the initial state $\rho_{in}=\rho_{N/D}$ as follows $(\bar{G}_E|G_{in})=\sum_{z{\in \mathcal{S}_{N/D}}} \bar G_E(z,z)$.
Thus, $S^z_0(t)$ requires the knowledge of the Liouvillian spectrum $E$ and of the amplitude $\alpha_{E}$ which encodes the initial conditions.

\begin{figure*}[ht!]
    \centering
\includegraphics[width=\linewidth]{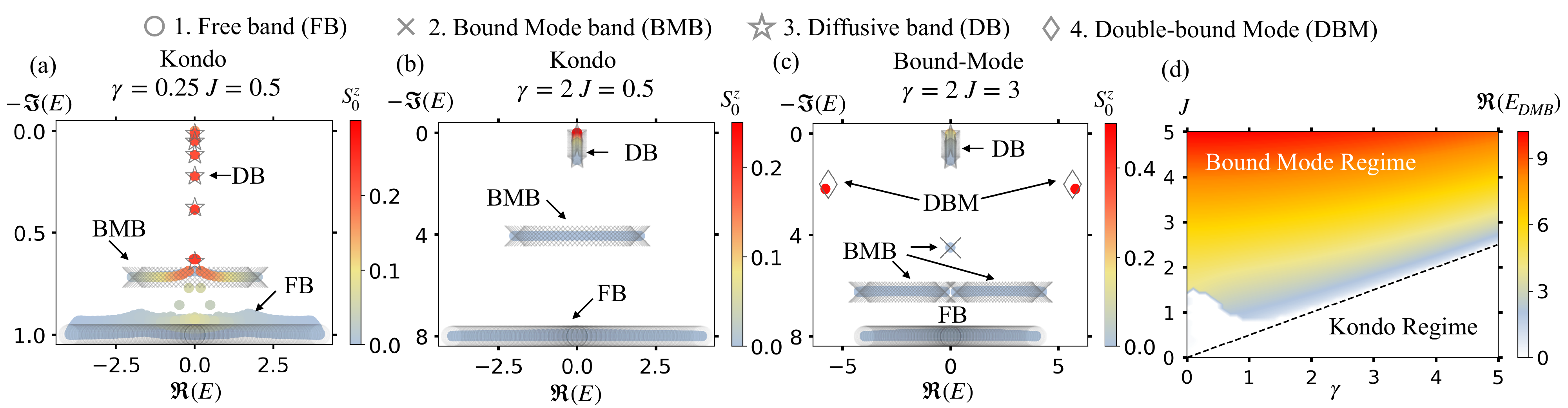}

    \caption{(a) The parameter space \((J, \gamma)\) is divided into two distinct regimes: the bound-mode regime and the Kondo regime. The real part of the double-bound-mode energy is computed for $L=30$ and is given in the heatmap. For large $\gamma \mathord{ \gg } 1$, the boundary between two regimes is almost $J\mathord{ = }\gamma/2$ as shown in the dashed line in (a). The Liouvillian spectrum and amplitude $\alpha_{E}= G_E(0,0)(\bar{G}_E|G_{in})$ for N\'eel state in the two-particle sector with dephasing rate $\gamma=0.25$ and impurity coupling $J=0.5$ (b) in the Kondo regime, $\gamma=2$,  $J=0.5$ (c) in the Kondo regime, and $\gamma=2$,  $J=3$ (d) in the bound-mode regime due to the existence of the double-bound mode.  The data points labeled in the legend with different shapes and numbers represent four different kinds (respectively correspond to the four kinds of solutions solved analytically in the main text). The scattering color plots are numerical results solved with system size $L=40$ where the heatmap shows absolute values of amplitude $\alpha_E^N$ for each eigenmode.  The modes with larger $\alpha_E^N$ dominate the impurity dynamics: the double-bound mode in the bound-mode regime (d) and the modes within the diffusive band in the Kondo regime (b, c). With a small noise $\gamma<1$, the diffusive band is mixed with the other bands (b), and when $\gamma>1$, the diffusive band is separated from the other bands (c,d).
    }
    \label{Eigenvalues}
\end{figure*}

\begin{figure}
    \hspace{-0.45cm}\includegraphics[width=0.95\linewidth]{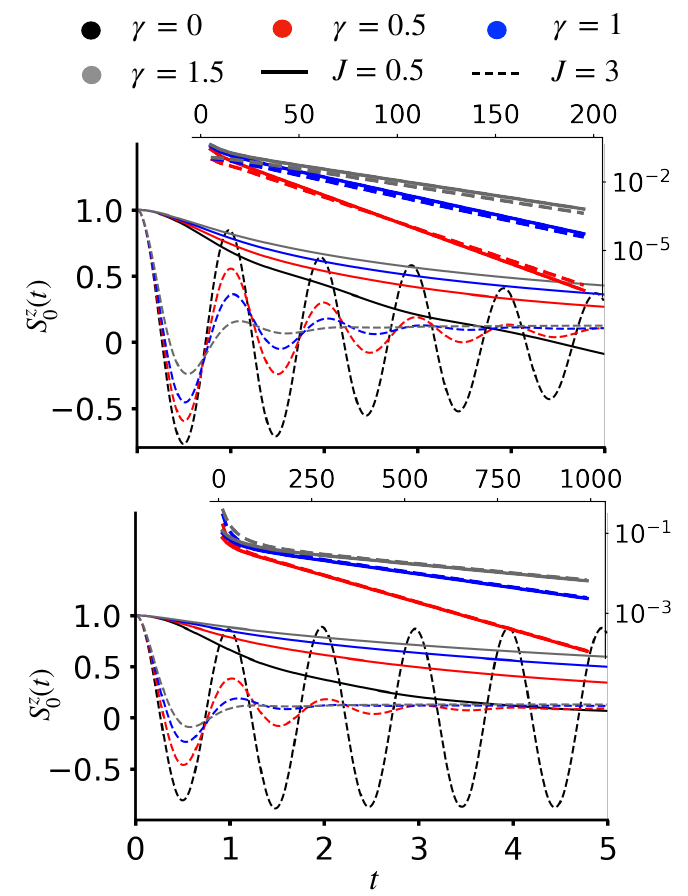}
    \caption{The dynamics of the XX  and XXX impurity models evolving from a N\'eel initial state under a Lindbladian evolution with and without dephasing noise. Different colors represent different dephasing rates $\gamma$, solid lines for $J=0.5$ in the Kondo regime, and dashed lines for $J=3$ in the bound-mode regime. The upper panel shows the data for the XXX chain for $L=10$ and the bottom panel for the XX chain for $L=40$. For both XX and XXX chains, in the bound mode phase, the signal decays faster as the dephasing rate increases. However, in the Kondo phase, the impurity decay is slower as the dephasing rate increases. 
    The insets in each panel show the data on a log-linear scale, displaying the same power law decay independent of $J$ in the long-time limit.
    }
    \label{Dynamics}
\end{figure}

We now proceed to solve for the spectrum $E$ of the Liouvillian $\hat{h}$ defined in Eq.~\eqref{fqham}.  
In the absence of the impurity,
this Liouvillian can be mapped to an integrable non-Hermitian Hubbard model after a unitary transformation~\cite{Essler-prozen,alba2023free,ziolkowska2020yang,prosen2008third}, here we generalize it to include the impurity as discussed in Appendix~\ref{mappingQC-Liouvillian}.
The Liouvillian in Eq.~\eqref{fqham} can thus be solved via the Bethe Ansatz (BA) \cite{faddeev1996algebraic,li2014exact,lieb1968absence}. It has  eigenvalue $E(k,q) = 2\cos k + 2\cos q -4i\gamma$ with the eigenmodes wavefunction taking the BA form
\begin{equation}
    G_{k,q}(x,y) = \frac{(-1)^y}{\mathcal{N}}\mathcal{S}\sum_{\sigma,\tau=\pm1}F_{\sigma k,\tau q}(x,y)e^{i\sigma kx+i\tau qy},
\end{equation}
where $F_{\sigma k,\tau q}(x,y) = A_{\sigma k,\tau q}\theta(y-x)+B_{\sigma k,\tau q}\theta(x-y)$, $\mathcal{N}$ is the normalization factor such that $|G_{{k,q}}|^2=1$ and $\mathcal{S}$ is the symmetrizer. The Liouvilian in Eq.\eqref{fqham} also has antisymmetrized eigenmodes~\cite{alba2023free}, but only symmetrized eigenmodes participate in dynamics since the initial configuration is symmetric, i.e. $G(x,y;0)=G(y,x;0)$. Here the amplitudes are related to the boundary scattering matrix $K(k)$ and the bulk scattering matrix $S(k,q)$ as: $A_{\sigma k,\tau q}=S(\sigma k,\tau q)B_{\sigma k,\tau q}$ and $A_{+k,\tau q}=K(k)A_{-k,\tau q}$~\cite{Essler-prozen,alba2023free,ziolkowska2020yang,karnaukhov2005exact} with
\begin{equation}
\begin{split}
    & K(k) = -\frac{e^{ik}+e^{-ik}-2i\gamma-J^2e^{ik}}{e^{ik}+e^{-ik}-2i\gamma-J^2e^{-ik}},
    \\& S(k,q) = \frac{\sin k - \sin q + 2\gamma}{\sin k - \sin q -2\gamma}.
\end{split}
\end{equation}
The quantization condition for quasimomenta $k$ and $q$ are obtained from the Bethe Ansatz equations (BAE)~\cite{karnaukhov2005exact}
\begin{equation}
    \begin{split}
    \label{BAE}
     e^{2ik(L+1)}&= K(k)S(k,q)S(-k,q),
     \\
     e^{2iq(L+1)}&= K(q)S(q,k)S(-q,k).
    \end{split}
\end{equation} 
If \((k, q)\) is a solution of the BAE~\eqref{BAE}, then \((\pm k +2 \pi, \pm q +2 \pi)\) is also a solution. This allows us to restrict the Brillouin zone to \(\Re k \in (0, \pi)\) with $\Re k$ ($\Im k$) as the real (imaginary) part of $k$. Furthermore, its complex conjugate \((k^*, q^*)\) also being a solution, indicates that if $E$ is a Liouvillian eigenvalue, then $-E^*$ is also an eigenvalue~\cite{alba2023free}.
The quasimomenta $(k,q)$ satisfying the BAEs fall into one of the following four cases in the limit $L\to \infty$. Here, we briefly summarize the results, and details can be found in Appendix~\ref{liouville spectrum}.

1.\textit{Free band}: When both S-matrices are regular \(S, K \neq 0,\infty\), the quasimomenta \(k\) and \(q\) are complex, with imaginary parts vanishing as \(\sim \frac{1}{L}\), \textit{i.e} \(k = \frac{\pi n_1}{L+1} + \mathcal{O}(L^{-1})\) and \(q = \frac{\pi n_2}{L+1} + \mathcal{O}(L^{-1})\), for \(n_1, n_2 = 1, \ldots, L\). These solutions exist for all parameters. These solutions dominate when $t\ll (4\gamma)^{-1}$ and generate noiseless dynamics.


2. \textit{Bound mode band:} When \(K \mathord{=} \infty\), quasimomenta are \(k_\pm = i\log\left(\frac{\pm\sqrt{J^2-1-\gamma^2}-i\gamma}{J^2-1}\right)\) and \(q = \frac{\pi n}{L+1} + \mathcal{O}(L^{-1})\) for \(n = 1, \ldots, L\). 
If \(\Im k_\pm < 0\), it is a bound mode; otherwise, it is not a valid solution. When both \(\Im k_\pm < 0\), another mode with \(E = 2\cos k_+ + 2\cos k_- - 4i\gamma\) exists. These solutions are similar as the free ones, but they generate bound-mode dynamics for early times.

3. \textit{Diffusive band:} When \(S \mathord{=} \infty\), the eigenspectra are \(E_n = 4i\sqrt{\gamma^2 - \sin\left(\frac{\pi n}{2L}\right)^2} - 4i\gamma + \mathcal{O}(L^{-3})\) with \(n = 0, 1, 2, \ldots, L\). Gapless excitations with \(E_n = -i\frac{\pi^2 n^2}{2\gamma L^2} + \mathcal{O}(L^{-3})\)~\cite{vznidarivc2015relaxation,Essler-prozen,alba2023free}, dominate the late time dynamics. These solutions with $\Re E_n= 0$ exist for all parameters, and they generate the diffusive dynamics due to the dephasing noise, known as Incoherentons\cite{haga2023quasiparticles}.


4. \textit{Double-bound mode:} There are two solutions where the particles are boundary-localized. When \(\gamma \mathord{<} 1\), the eigenvalues can be solved perturbtively \(E_\pm = 4\cos k_\pm + 4i\gamma \left[\sum_{x=1}^L \mathcal{N}^2 e^{i4k_\pm x} + \frac{\mathcal{N}}{J^2} - 1\right] + \mathcal{O}(\gamma^2)\), where \(k_\pm = \frac{i}{2}\log(J^2 - 1) + \pi/2 \pm \pi/2\) and \(\mathcal{N} = \frac{1}{J^2} + \frac{e^{-2ik_\pm}}{1-e^{-2ik_\pm}}\). When \(\gamma > 1\), the eigenvalues solved with an effective two-site Liouvillian are \(E_\pm = \pm\sqrt{4J^2 - \gamma^2} - i\gamma\). These solutions dominate the early-time oscillation in the magnetization in the bound-mode regime at short times.

Fig.~\ref{Eigenvalues} shows the spectrum of the Liouvillian in the Kondo and bound regimes. We present the cross-over diagram that delineates these two regimes in Fig.~\ref{Eigenvalues}(a) through the value of the real part of the Double bound mode $\Re E_{DBM}$. A non-zero $\Re E_{DBM}$ signifies the bound-mode regime, while a zero value indicates the Kondo regime.
We further analyze the spectrum's real and imaginary parts in the Kondo regime in Fig.~\ref{Eigenvalues}(a) and (b), as well as the bound-mode regime in Fig.~\ref{Eigenvalues} (c). We see the free band (spectrum 1; circles), the bound mode band (spectrum 2; crosses), and the diffusive band (spectrum 3; star) in the Kondo regime, where the majority of the weight in $\alpha_{E}^N$ lies in the diffusive band. In the bound-mode regime, we also see the appearance of double bound modes (spectrum 4; diamonds) that have the largest weight in the coefficients $\alpha_{E}^N$ and 
hence they dominate the early-time impurity dynamics $S^z_0(t)$.

In~\cite{kattel2024kondo}, we showed that at the XX point without noise, there is a boundary phase transition at $J=\sqrt{2}$. When $J<\sqrt{2}$, the impurity is screened by the many-body Kondo effect, whereas when $J>\sqrt{2}$, the impurity is screened by single particle bound mode localized at the edge. Here, we turn to study the dynamical characterization of the boundary phases. In the bound mode phase ($J>\sqrt{2}$), the time-dependent magnetization shows oscillatory behavior $S^z_0(t)\sim \cos(\frac{J^2}{\sqrt{J^2-1}}t)$ when $t\to\infty$. However, in the Kondo phase ($J<\sqrt{2}$), the impurity magnetization leaks to the bulk with the time scale proportional to the Kondo time $t_K=\sqrt{|J^2-1|}/J^2$.  Detailed calculation showing this sharp phase transition in dynamical quantities is in Appendix~\ref{noiseless-quench}.

When noise is introduced ($\gamma > 0$), there is no longer a sharp phase transition. Instead, the impurity magnetization $S^z_0(t)$ decays similarly in both regimes at late times, exhibiting the Zeno effect, where the dynamics slows down as noise strength increases. This behavior is driven by the diffusive band as it has the largest overlap $\alpha_E^N$. The modes in the diffusive band have $|\Im E(\gamma)|\propto \gamma^{-1}$, so the dynamics slows down as $\gamma$ increases and exhibits the quantum Zeno effect. In contrast, the early-time dynamics differ between the two regimes. In the \textit{bound-mode regime}, the impurity dynamics is dominated by the double-bound mode, which has the largest overlap $\alpha_E^N$ and $|\Im E(\gamma)|\propto \gamma$. Thus, the impurity decays faster as $\gamma$ increases. However, in the \textit{Kondo regime}, the early-time dynamics are also influenced by the diffusive band—similar to the long-time limit where the decay rate decreases with increasing noise, resulting in the quantum Zeno effect. The model parameters $J$ and $\gamma$ corresponding to these two regimes are shown in Fig.~\ref{Eigenvalues}(a). 

Numerical results for $S^z_0(t)$ starting from the N\'eel state at the XX point are shown in the bottom panel of Fig.~\ref{Dynamics} with two situations when the system is in the Kondo regime $J=0.5$ and in the bound-mode regime $J=3$.
As seen from the data, for the early time in the bound-mode regime, the noise adds a decaying envelope that is accelerated as the dephasing rate is increased. This is an expected result because the boundary mode is essentially a two-qubit state, and qubits subjected to dephasing noise decay faster as the decay rate increases. In the Kondo regime or after the double bound mode decays \textit{i.e.} $t\sim \gamma^{-1}$, the noise slows the leakage of magnetization and leads to a quantum Zeno effect as the dephasing rate increases. In the long-time limit, independent of the boundary coupling $J$, the impurity magnetization falls off as $S^z_0(t)\sim e^{-t/\tau}$ with $\tau\propto \frac{1}{\gamma L^2}$ in both regimes. This behavior of the impurity in the Kondo regime is consistent with the heating of the bulk induced by Hermitian Lindblad operators~\cite{HLPhysRevB.93.094205,HLPhysRevLett.116.160401,HLPhysRevLett.116.237203}, driving the impurity to (asymptotically free) weak coupling behavior~\cite{hewson1997kondo,kondo2012physics,andrei1983solution,tsvelick1983exact}.


\begin{figure}
    \centering
\includegraphics[width=0.9\linewidth]{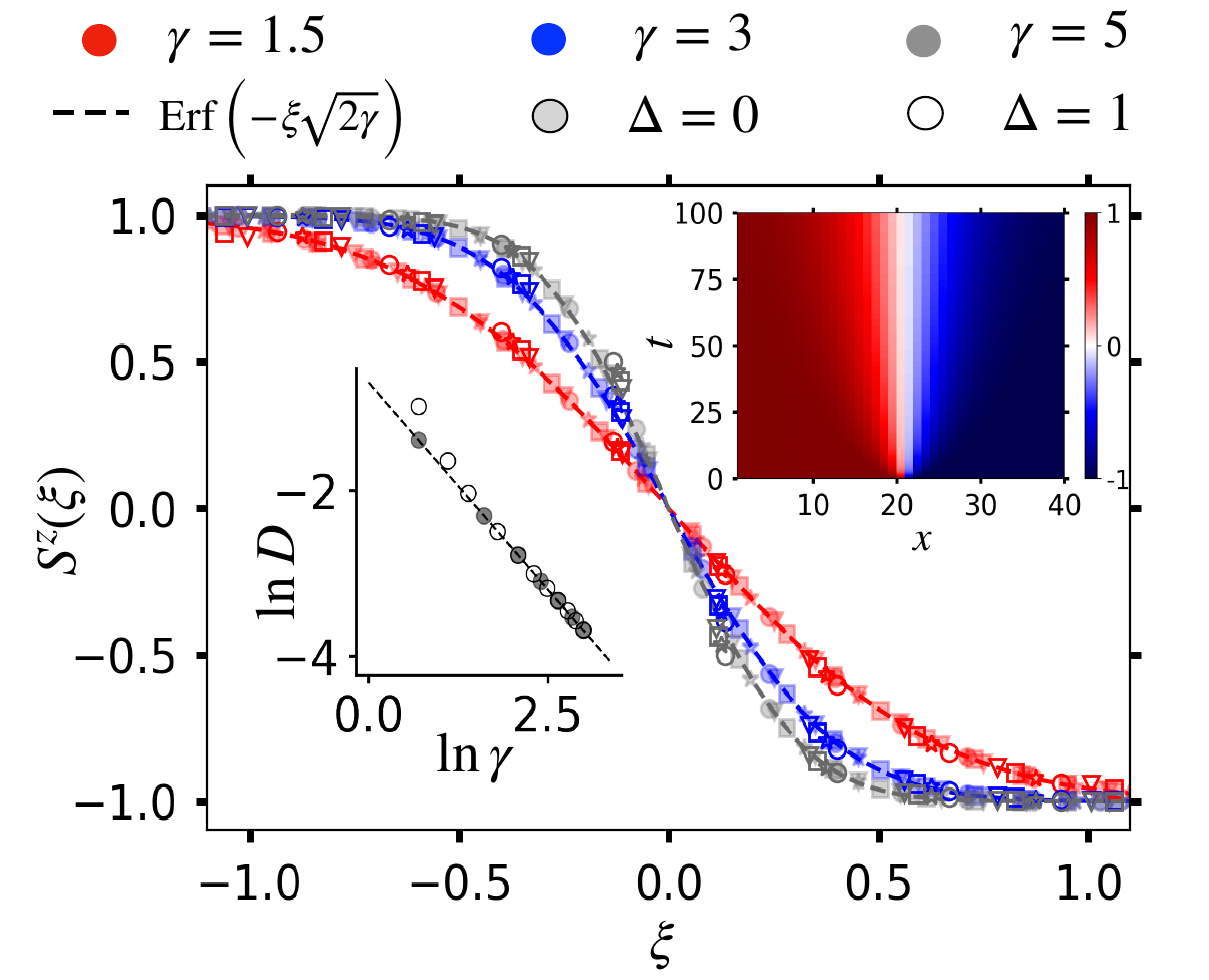}
    \caption{Data collapse for $S^z(x,t) =S^z(\xi)$ with $\xi = xt^{-\frac{1}{2}}$ in the XX case (filled markers, with system size $L=40$) and XXX case  (hollow markers, with system size $L=10$) starting from a domain wall initial state (as shown in the color plot in the right inset for XX case with $\gamma=2$) under various dephasing rates $\gamma$. Different marker shapes correspond to the spin configuration at different time slices, and different marker color refers to different dephasing rates. The left inset shows the dependence of the diffusion constant $D$ on $\gamma$ (filled markers for XX case and hollow for XXX case), and the dashed lines are the analytical result for the noisy XX chain $D = \frac{1}{2\gamma}$.
    }
    \label{S(x,t)}
\end{figure}

Although no noise is applied at the impurity site, the Zeno effect on impurity is present in the Kondo regime at all times and at late times ($t>\gamma^{-1}$) in the bound mode phase, because noise changes the bulk transport property from ballistic when the system is at the XX point to diffusive~\cite{jin2020generic} with diffusion constant $D=\frac{1}{2\gamma}$ as shown in \cite{alba2023free}.

\textit{Heisenberg (XXX) chain:} Next, we examine the Trotterized dynamics of the interacting circuit corresponding to the isotropic Heisenberg model ($\Delta = 1$) with a boundary impurity. The equilibrium boundary phases have been studied in~\cite{wang1997exact,kattel2023kondo,frahm1997open}. We show here that the bound mode and Kondo phases also exist dynamically, manifesting themselves in a similar way as in the XX case: the bound mode phase exists when $J>4/3$ where the impurity magnetization oscillates in time, and the Kondo phase when $J<4/3$ where it decays as shown in Fig.~\ref{Dynamics} in the noiseless case.   The bulk transport properties in the noiseless case is compatible with the KPZ universality class as shown in~\cite{ljubotina2019kardar,rosenberg2024dynamics}.  

Turning on the noise, we find that it impacts both the bulk and boundary dynamics similarly to the XX case discussed above. 
To show the bulk dynamics is diffusive for XXX case, we prepare the system in a domain wall state where all sites on the left (right) half are up (down) and compute the magnetization $S^z_j(t)$ as time evolves.  When the system is diffusive,  in the continuum limit as lattice spacing $a\to 0$ and system size $L\to\infty$,  the magnetization $S^z_j(t)\to S^z(x,t)$ satisfies the diffusion equation i.e., $(\partial_t-D\partial_x^2)S^z(x,t)=0$, with initial condition $S^z(x,t\mathord{=}0)=\mathrm{Sign}(L/2-x)$.
The numerical results for the dynamics of both XX and XXX cases are shown in Fig.~\ref{S(x,t)}, where all the data collapses on the curve $S^z(\xi) = -\mathrm{Erf}(\xi/\sqrt{D})$ with $\xi = x/\sqrt{t}$ which is the solution of diffusion equation from the domain wall initial configuration. For large values of $\gamma$, we find clear numerical evidence that $D=\frac{1}{2\gamma}$ as in the XX case, but for smaller values of $\gamma$, it deviates (as shown in the left inset in Fig.~\ref{S(x,t)}) due to finite size and time effects (see Appendix~\ref{nummmdets}) and we therefore expect that the relation $D=\frac{1}{2\gamma}$ holds for each value of $\gamma$ considered. Some of these results were obtained in ~\cite{esposito2005emergence,eisler2011crossover,vznidarivc2015relaxation,vznidarivc2017dephasing}.   
Here we focus on the fact that the bulk diffusive dynamics reflects itself in the impurity dynamics as shown in the impurity magnetization dynamics from the initial N\'eel state with and without noise in the upper panel of Fig.~\ref{Dynamics}. 
In the presence of noise, both Kondo and bound-mode regimes exist, and the impurity in each regime shows a similar behavior as in the free XX chain. In the bound-mode regime, the impurity magnetization initially decays faster as the noise strength is increased, and after the bound modes vanish ($t>\gamma^{-1}$), the impurity experiences the quantum Zeno effect, whereas in the Kondo regime, the impurity magnetization exhibits the Zeno effect in all time scales.

\textit{Conclusion:} Using spin chain models, we show how dissipation alters unitary dynamics differently in different phases: dephasing noise suppresses impurity magnetization decay in the Kondo regime (quantum Zeno effect) but accelerates initial decay in the bound-mode regime before Zeno dominance. In the long-time limit, both regimes converge to diffusion-driven Zeno dynamics due to the bulk transport becoming diffusive when the dephasing noise is introduced ~\cite{esposito2005emergence,eisler2011crossover,vznidarivc2015relaxation,vznidarivc2017dephasing}.


Acknowledgment: We thank T.~Giamarchi, V.~Alba, F.~Essler, and D.~Huse for useful discussions. J.H.P. is partially supported by the Army Research Office Grant No.~W911NF-23-1-0144 and the Office of Naval Research grant No.~N00014-23-1-2357 (J.H.P.).

\bibliography{ref}

\onecolumngrid 

\newpage

\begin{appendix}

In the following appendices, we first map the circuit dynamics to the Lindbladian evolution. Then, at the XX point, we show that the corresponding Lindbladian can be mapped to an integrable Hubbard Hamiltonian with imaginary onsite interaction and boundary hopping defects in doubled Liouville-Fock space. We then solve the dynamics of the noiseless model with the free-fermion technique. After that, we solve the non-Hermition Hubbard model in the two-particle sector to discuss the dynamics in the presence of noise. We also discuss the derivation of the Bethe-Ansatz equations for the complex onsite Hubbard model in an arbitrary particle sector. In the final section, we show the details of numerical results.

\section{Mapping from Quantum circuits to Liouvillian dynamics}
\label{mappingQC-Liouvillian}

Consider the trigonometric six vertex R-matrix \cite{baxter2007exactly,sutherland2004beautiful}
\begin{equation}
   R(u)=\left(
\begin{array}{cccc}
 1 & 0 & 0 & 0 \\
 0 & \frac{\sinh (u)}{\sinh (u+\eta)} & \frac{\sinh (\eta)}{\sinh (u+\eta)} & 0 \\
 0 & \frac{\sinh (\eta )}{\sinh (u+\eta)} & \frac{\sinh (u)}{\sinh (u+\eta)} & 0 \\
 0 & 0 & 0 & 1 \\
\end{array}
\right)
\end{equation}

    which is  a solution of the Yang-Baxter equation~\cite{faddeev1996algebraic}
\begin{equation}
    R_{12}(u-v)R_{13}(u)R_{23}(v)=R_{23}(v)R_{13}(u)R_{12}(u-v).
\end{equation}
Notice that the Yang-Baxter equation remains satisfied if we shift $u_i\to u_i-\theta_i$ where $\theta_i$ are arbitrary inhomogeneous parameters.

   Let us consider the following two single-row transfer matrices
 \begin{align}
T_A(u) &= R_{A,L}(u-\theta_{L})R_{A,L-1}(u-\theta_{L-1})\cdots R_{A,2}(u-\theta_2)R_{A,1}(u-\theta_1)\nonumber\\
\hat{T}_A(u) &= R_{A,1}(u+\theta_1)R_{A,2}(u+\theta_2)\cdots R_{A,L-1}(u+\theta_{L-1})R_{A,L}(u+\theta_{L}),
\end{align}
where $1,2,\ldots,L$ are the labels for the physical space and $A$ is an auxiliary space.

Furthermore, denoting the physical space of impurity as $0$, we define
\begin{align}
    K_A(u)=R_{A,0}(u-\theta_0-d)R_{A,0}(u+\theta_0+d),
\end{align}
such that the $K$ matrices satisfy the reflection equations
\begin{equation}
    R_{ij}(\lambda - \mu) K_{iA}(\lambda) R_{ji}(\lambda + \mu) K_{jA}(\mu) = K_{jA}(\mu) R_{ij}(\lambda + \mu) K_{iA}(\lambda) R_{ji}(\lambda - \mu).
\end{equation}

Now, we define the monodromy matrix
\begin{equation}
    \Xi(u)=T_A(u)K_A(u)\hat T_A(u).
\end{equation}
The trace of the monodromy matrix over the auxiliary space is defined as the double-row transfer matrix~\cite{wang2015off}
\begin{equation}
    t(u)=\operatorname{tr}_A\Xi(u).
    \label{tmat}
\end{equation}
    
We fix the inhomogeneity parameters~\cite{ljubotina2019ballistic}
\begin{equation}
    \theta_0=\theta_2=\theta_4=\ldots=\theta_L=\frac{\lambda}{2}\quad\quad \theta_1=\theta_3=\theta_5=\ldots=\theta_{L-1}=-\frac{\lambda}{2}.
\end{equation}

Such that when the transfer matrix Eq.\eqref{tmat} is evaluated at $u=\lambda$, we obtain
\begin{equation}
   \mathbb{U}= t(k)=\mathcal{U}_{0,1}U_{2,3}U_{4,5}\ldots U_{L-1,L}U_{1,2}U_{3,4}U_{5,6}\ldots U_{L-2,L},
\end{equation}

where
\begin{align}
    U_{j,j+1}&=\check R_{j,j+1}(\lambda)=P_{j,j+1}R_{j,j+1}(\lambda),\nonumber\\
    \mathcal{U}_{0,1}&=R_{0,1}(-d)R_{0,1}(\lambda+d)=\check R_{0,1}(-d)\check R_{0,1}(\lambda+d).
\end{align}
Here $P_{ij}=\frac{1}{2}(\mathbb{I}_{i,j}+\sigma^x_i\sigma^x_j+\sigma^y_i\sigma^y_j+\sigma^z_i\sigma^z_j)$ is the permutation matrix and explicit form of the $\check R$ matrix reads
\begin{align}
    \check R^\eta_{i,j}(\lambda)=\frac{1}{2} \left( \mathbb{I}_{i,j} + \sigma^z_i \sigma^z_j \right) 
+  \frac{\sinh(\lambda)}{2\sinh(\eta + \lambda)} \left( \sigma^x_i \sigma^x_j + \sigma^y_i \sigma^y_j \right)+  \frac{\sinh(\eta)}{2\sinh(\lambda + \eta)} \left( \mathbb{I}_{i,j} - \sigma^z_i \sigma^z_j \right).
\end{align}


These local unitary can be written as 
\begin{align}
    U_{j,j+1}&=e^{-i \frac{J}{2} \delta t (\sigma^x_j \sigma^x_{j+1}+\sigma^y_j \sigma^y_{j+1}+\Delta (\sigma^z_j \sigma^z_{j+1}-1))},\\
    \mathcal U_{0,1}&=e^{-i \frac{J'}{2} \delta t (\sigma^x_j \sigma^x_{j+1}+\sigma^y_j \sigma^y_{j+1}+\Delta' (\sigma^z_j \sigma^z_{j+1}-1))}
\end{align}

upon identifying
\begin{equation}
    e^{ i J \delta t (\Delta \mp 1) }=\frac{e^{\eta }\pm e^{ \lambda}}{e^{ \lambda+\eta }\pm 1}\quad\quad \text{and}\quad\quad e^{ i J' \delta t (\Delta' \mp 1) }=\frac{\left(e^{d+\eta }\pm 1\right) \left(e^{d+ \lambda}\pm e^{\eta }\right)}{\left(e^d\pm e^{\eta }\right) \left(e^{d+ \lambda+\eta }\pm 1\right)}.
\end{equation}


The circuit in Fig.~\ref{IntegrableCircut} corresponds to the unitary matrix of the following form
\begin{equation}
\begin{split}
    &\mathbb{U}(\delta t) =  \mathcal{U}_{0,1}\prod_{j\in 2\mathbb{Z}}U_{j,j+1}\prod_{j\in 2\mathbb{Z}-1} U_{j,j+1}= 1-iH\delta t+\mathcal{O}(\delta t^2),
\end{split}
\end{equation}
where in the Trotterization limit, we obtain an integrable Hamiltonian with boundary impurity of the form
\begin{align}
    H=\sum_{i=1}^{L-1}\frac{J}{2}(\sigma_i^x \sigma_{i+1}^x+\sigma_i^y \sigma_{i+1}^y+\Delta(\sigma_i^z \sigma_{i+1}^z-1))+\frac{J'}{2} \left(\sigma_0^x \sigma_{1}^x+\sigma_0^y \sigma_{1}^y+\Delta' (\sigma_0^z \sigma_{1}^z-1)\right),
\end{align}
where the parameters in two representations are related by $\Delta=\cosh(\eta), \Delta'=\frac{\cosh(\eta)}{\cosh( d)}$ and $J'=J \frac{\sinh^2(\eta)\cosh(d)}{\sinh^2(\eta)-\sinh^2(d)}$.

Notice that both $J'$ and $\Delta'$ are expressed through a single free parameter $d$ hence the model is not integrable for arbitrary $\Delta'$ and $J'$. However, at the isotropic $\eta=0$~\cite{kattel2023kondo,wang1997exact} and XX $\eta=i\frac{\pi}{2}$~\cite{kattel2024kondo} limit, the boundary and bulk anisotropy parameter becomes equal \textit{i.e.} $\Delta'=\Delta$ and $J'$ becomes an arbitrary free parameter.

Taking into account the presence of noise while applying the gates, the density matrix of the qubits can be described by a discrete-time evolution as
\begin{equation}
    \rho(t+\delta t) = \sum_{j} M^\dagger_j\rho(t)M_j,
\end{equation}
with $M_0 = 1-\Delta t (K+iH)$, for $j\neq 0 $, $M_j = \sqrt{\delta t}L_j$ and  $K = \frac{1}{2}\sum_j L^\dagger_jL_j \delta t$  and $L_j$ are the Lindbladian jump operators. Here, we will concentrate solely on the case where $L_j = \sigma^z_j$. Taking $\delta t \to 0$, we obtain the Lindbladian-master equation~\cite{breuer2002theory}
\begin{equation}
    \frac{d}{dt}\rho(t) = \mathcal{L}(\rho),
\end{equation}
where the Liouvillian operator can be written as 
\begin{equation}
\label{LeqS}
    \mathcal{L}(\rho) = -i\left[H,\rho\right] + \gamma\sum_i \left[L^\dagger_i \rho L_i-\frac{1}{2}\left\{L^\dagger_i L_i,\rho\right\} \right]
\end{equation}
with $\{A,B\}=AB+BA$ and $[A,B]=AB-BA$ being the anticommutator and  the commutator, respectively.
\section{XX chain point of the Hamiltonian}\label{hubbardmap}
At the XX chain point \textit{i.e} when $\eta = i\pi/2$, we end up with a Hamiltonian of a XX chain with an XX impurity~\cite{kattel2024kondo}
\begin{equation}
    H = \sum_{i=1}^{L-1}\frac{1}{2}\left(\sigma^x_i\sigma^x_{i+1}+\sigma^y_i\sigma^y_{i+1}\right)+\frac{J}{2}\left(\sigma^x_0\sigma^x_{1}+\sigma^y_0\sigma^y_{1}\right).
\end{equation}
 After performing the Jordan-Wigner transformation, one can write the model in terms of Fermionic operators i.e. the Hamiltonian becomes
\begin{equation}
    H = \sum_{i=0}^{L-1}\left(c^\dagger_ic_{i+1}+c^\dagger_{i+1}c_i\right)+J\left(c^\dagger_0c_{1}+c^\dagger_{1}c_0\right),
\end{equation}
and the Liouvillian operator takes the form
\begin{equation}
    \mathcal{L}(\rho) = -i\left[H,\rho\right] + 4\gamma \sum_{i=1}^{L-1} \left(c^\dagger_{i}c_i \rho c^\dagger_{i}c_i-\frac{1}{2}\{\rho,c^\dagger_ic_i\}\right).
\end{equation}
The density matrix $\rho$ is an operator defined in the $2^N$ dimensional Hilbert space, which can be written as $\rho = \sum_{m,n} \rho_{mn}\ket{m}\bra{n}$. We can rewrite it as a vector in a $4^N$ dimensional Hilbert space, so-called the Liouville-Fock space, as $\ket{\rho} = \sum_{m,n}\rho_{mn}\ket{n,m}\rangle$ without losing information~\cite{prosen2008third}. Then, the Liouvillian operator acts as an operator in the enlarged vector space in the following way
\begin{equation}
    \mathcal{L}\ket{\rho}\rangle = \frac{d}{dt}\ket{\rho}\rangle,
\end{equation}
with the Liouvillian as superoperator
\begin{equation}
    \mathcal{L} = -iH +i\Tilde{H} + 4\gamma \sum_{i=1}^{L-1}\left(c^\dagger_ic_i\Tilde{c}^\dagger_i\Tilde{c}_i-\frac{1}{2}c^\dagger_ic_i-\frac{1}{2}\Tilde{c}^\dagger_i\Tilde{c}_i\right).
\end{equation}
Here, we introduce a new superoperator acting on the Liouville-Fock space in the following way:
\begin{equation}
\begin{split}
    c^\dagger_i\ket{n,m}\rangle &= c^\dagger_i\ket{n}\bra{m}
    \\\Tilde{c}^\dagger_i\ket{n,m}\rangle &= \ket{n}\bra{m}c_i
\end{split}
\end{equation}
If we rename $c^\dagger_i = c^\dagger_{i,\uparrow}$ and $\Tilde{c}^\dagger_i = c^\dagger_{i,\downarrow}$, and perform the unitary transformation \begin{equation}
\label{Utrans}
    U=\prod_{i=1}^{N/2}(1-2\Tilde{c}^\dagger_{2i}\Tilde{c}_{2i}),
\end{equation} the Liouvillian operator becomes the Hubbard model with imaginary coupling \cite{Essler-prozen}
\begin{equation}
    \mathcal{L}=-iJ(c^\dagger_{0,\sigma}c_{1,\sigma}+c^\dagger_{1,\sigma}c_{0,\sigma})-i\sum_{i=1}^{L-1}(c^\dagger_{i,\sigma}c_{i+1,\sigma}+c^\dagger_{i+1,\sigma}c_{i,\sigma})+\gamma\sum_{i=1}^{L}(2c^\dagger_{i,\uparrow}c_{i,\uparrow}-1)(2c^\dagger_{i,\downarrow}c_{i,\downarrow}-1)-\gamma (L-1),
\end{equation}

\begin{equation}
    \mathcal{L}=J(c^\dagger_{0,\sigma}c_{1,\sigma}+c^\dagger_{1,\sigma}c_{0,\sigma})+\sum_{i=1}^{L-1}(c^\dagger_{i,\sigma}c_{i+1,\sigma}+c^\dagger_{i+1,\sigma}c_{i,\sigma})+i\gamma\sum_{i=1}^{L}(2c^\dagger_{i,\uparrow}c_{i,\uparrow}-1)(2c^\dagger_{i,\downarrow}c_{i,\downarrow}-1)-i\gamma (L-1),
\end{equation}
where $\{c^\dagger_{i,\sigma},c_{j,\sigma'}\}=\delta_{ij}\delta_{\sigma\sigma'}$, and all other combinations anti-commute. In this formalism, the Lindbladian-master equation becomes an eigenvalue problem
\begin{equation}
    \label{Lrho=Erho}\mathcal{L}\ket{\rho_n}\rangle=E_n\ket{\rho_n}\rangle.
\end{equation}
One thing worthy of mentioning is that the dynamics of 2$k$-point correlation function $G^{x_1,...,x_k}_{y_1,...,y_k}(t) = \text{Tr}\left\{c^\dagger_{x_1}...c^\dagger_{x_k}c_{y_1}...c_{y_k}\rho(t)\right\}$, under this mapping, can be shown to be governed by the 2$k$-particle Schrodinger equation~\cite{Essler-prozen}. Particularly for the two-point correlation function we are interested in, 
\begin{equation}
\label{S2ptcorr}
  i\frac{d}{dt}G(x,y,t) = \left[J_{x}\Delta_x+J_{y}\Delta_y+4i\gamma\delta_{x,y}+2i\gamma(\delta_{x}+\delta_{y}-2)\right]G(x,y,t),
\end{equation}
 where we introduced a compact notation $J_x\mathord{=} J$ when $x\mathord{=}0$, $J_x\mathord{=}1$ otherwise and $\Delta f(x)=f(x-1)+f(x+1)$.
\section{Noiseless XX-impurity Quench}\label{noiseless-quench}
\begin{figure}
    \centering
    \includegraphics[width=0.8\textwidth]{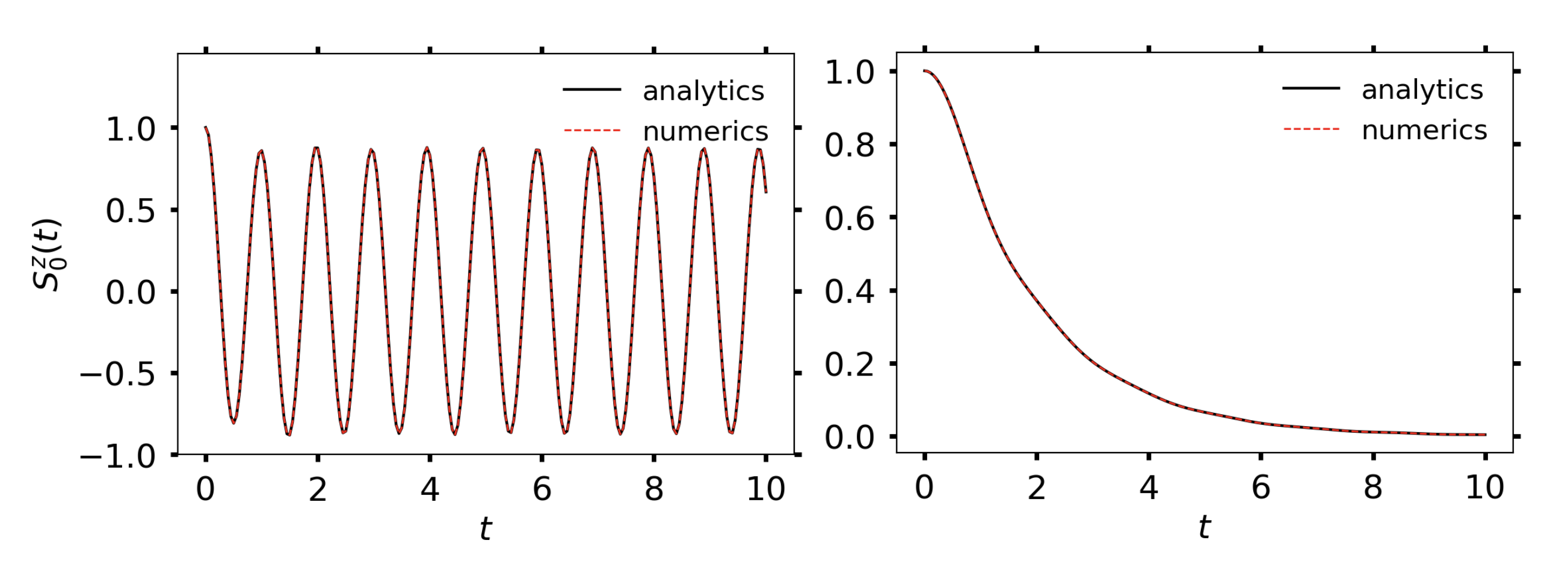}
    \caption{Dynamics of the spin at the impurity site with initial N\'eel state. The system is in bound mode phase $J=3$ (left panel) where the analytic curve of the time-dependent spin profile is given by Eq.~\eqref{mageqn} and numerical cure is calculated for $L=40$ with exact diagonalization. The system is in the Kondo phase with $J=0.5$ (right panel) where the analytic curve of the time-dependent spin profile is given by Eq.~\eqref{mageqn} and numerical cure is calculated for $L=40$ with exact diagonalization.}
    \label{fig:dynamics}
\end{figure}

Let us first discuss the dynamics of the model in the absence of the noise. We consider the quench dynamics of the free model governed by a quantum mechanical Hamiltonian
\begin{equation}
    h = \sum_{j=0}^{L-1}J_j\Delta_j,
\end{equation}
where $J_x\mathord{=} J$ when $x\mathord{=}0$, $J_x\mathord{=}1$ other wise and $\Delta_jf(j) = f(j+1) + f(j-1)$ is the discrete derivative. This first quantized Hamiltonian has eigenfunction
\begin{equation}
    F_k(j) = \frac{1}{J_j}(Ae^{ikj}+Be^{-ikj}).
\end{equation}
The boundary condition when $x=0$ gives ~\cite{kattel2024kondo}
\begin{equation}
\label{Sboundary}
    S_b(k) = \frac{B}{A} = -\frac{(J^2-1)e^{ik}-e^{-ik}}{(J^2-1)e^{-ik}-e^{ik}}
\end{equation}
and $x=L$ gives the quantization condition of $k$
\begin{equation}
   \frac{(J^2-1)e^{ik}-e^{-ik}}{(J^2-1)e^{-ik}-e^{ik}} = e^{i2(L+1)k}.
\end{equation}
There exists a bound mode with $\Im k >0$, described by $B=0$ such that the wave function is normalizable in the thermodynamic limit, i.e. $(J^2-1)e^{ik}-e^{-ik}=0$ which gives $k_{BM} = \frac{i}{2}\log(J^2-1)$ with bound mode energy $E =2\cos k= \frac{J^2}{\sqrt{J^2-1}}$ and $\Im k >0$ gives $J>\sqrt{2}$. Likewise, $A=0$ gives $k_{BM} = -\frac{i}{2}\log(J^2-1)$.
After studying the spectrum and equilibrium physics, we can now study the quench dynamics of this problem. Since the model is non-interacting, we need to solve only the single-mode evolution and take the algebraic sum of each mode at the end of the calculation. 
We prepare the initial state as $F(j,t=0) = F_0(j) = \delta_{j,j_0}$, and we solve for $F(j,t)$. The dynamics of a wave function is given by the following equation
\begin{equation}
\begin{split}
    F(x,t)&=\sum_{\ell,k} F^*_k(\ell)F_0(\ell)F_k(j)e^{i\epsilon_k t}\\
    & = \sum_{k} F^*_k(j_0)F_k(j)e^{i\epsilon_k t}.
\end{split}
\end{equation}
Let us look at the simplest case where we prepare the fermion (spin flip)  at the impurity site initial $j_0=0$ and ask how the fermion number (magnetization) changes as time evolves; in the Kondo phase $J<\sqrt{2}$
\begin{equation}
    n_0(t) = F(0,t)^2 = \left(\int \frac{dk}{\pi} \frac{J^2\sin^2(k)}{J^4-4(J^2-1)\cos^2(k)} e^{-i2t\cos k}\right)^2,
\end{equation}
with $|F_k(0)|^2 = \frac{1}{J^2}(A+B)^2 = \frac{J^2\sin^2(k)}{J^4-4(J^2-1)\cos^2(k)}$.
Changing the variable $\epsilon = 2\cos k$, we obtain
\begin{equation}
\label{Fkondo}
    F_{Kondo}(0,t)  = \frac{1}{2J^2\pi}\int_{-2}^{2} d\epsilon \frac{\sqrt{4-\epsilon^2}}{1-\mathrm{Sign}(J^2-1)t_K^2\epsilon^2} e^{-i\epsilon t},
\end{equation}
where the Kondo time $t_K = \frac{\sqrt{|J^2-1|}}{J^2}$ is the only scale in this integral. We verify this result numerically as shown in Fig.\ref{fig:dynamics}.
In the Bound Mode Phase, the same quantity can be expressed as the following 
\begin{equation}
\label{Fbm}
    F_{BM}(0,t)  = \frac{1}{2J^2\pi}\int_{-2}^{2} d\epsilon \frac{\sqrt{4-\epsilon^2}}{1-\mathrm{Sign}(J^2-1)t_K^2\epsilon^2} e^{-i\epsilon t} + \frac{J^2-2}{J^2-1}\cos(E_b t).
\end{equation}
The last cosine term is from the bound mode with energy $E_b = \pm \frac{J^2}{\sqrt{J^2-1}}$.

In the Kondo phase, the wavefunction at the $\ell$ spin, if we flip $j$ initially, is given by
\begin{equation}
    F_K(j,\ell,t)  = \int_0^{\pi} \frac{dk}{2\pi} \frac{1}{J_\ell J_j}(e^{-ikj}+S_b(k)e^{ikj})(e^{ik\ell}+S_b(k)^*e^{-ik\ell})e^{-i2t\cos k}.
\end{equation}
In the bound mode phase,
\begin{equation}
    F_B(j,\ell,t) = F_K(j,\ell,t) + \frac{J^2}{J_jJ_\ell}\frac{J^2-2}{J^2-1}\left[\frac{e^{-iE_Bt}}{(\sqrt{J^2-1})^{j+\ell}}+\frac{e^{iE_Bt}}{(-\sqrt{J^2-1})^{j+\ell}}\right].
\end{equation}
The magnetization at site $\ell$ with $\{j_x\}$ spin initially flipped is given by
\begin{equation}
    S^z(\ell,t) = 2\sum_{j\in\{j_x\}}|F(j,\ell,t)|^2-1.
    \label{mageqn}
\end{equation}
with $F=F_K$ or $F=F_B$ depending on which phase the system is in.

\section{Liouvillian Spectrum}\label{liouville spectrum}
In the previous section, we solved XX-impurity Model quenched from an initial state with only some spins flipped, i.e. $\ket{\psi_0} = \prod_{j_x}\sigma^+_{j_x}|\downarrow,\downarrow,...\downarrow\rangle$. Now, we want to ask how noise affects this model. Using Eq.~\eqref{S2ptcorr}, we need to study the two-particle sector of this model to solve for the dynamics of the two-point correlation function.

In the Liouville-Fock space, our initial state (initial condition of the correlation matrix) with spin-flip at site $j_x$ corresponds to the initial density matrix $\ket{\rho_0}\rangle = c^\dagger_{j_x\uparrow}c^\dagger_{j_x\downarrow}\ket{0}\rangle$. If we initially flip more spins, we just need to superpose them, i.e. $\ket{\rho_0}\rangle = \sum_{x_j}c^\dagger_{x_j\uparrow}c^\dagger_{x_j\downarrow}\ket{0}\rangle$. Due to the $U(1)\times U(1)$ symmetry of the Liouvillian operator, the number of fermions and total magnetization are conserved separately. Thus, we only need to diagonalize the eigenvalue problem in Eq.~\eqref{Lrho=Erho} in two particles $N=2$  and one spin-flip $M=1$ sector.

The time-independent Schrödinger equation read from Eq.\eqref{S2ptcorr} is
\begin{equation}
EG(x,y) = \left[J_{x}\Delta_x+J_{y}\Delta_y+4i\gamma\delta_{x,y}+2i\gamma(\delta_{x}+\delta_{y}-2)\right]G(x,y),
\end{equation}
with eigenvector
\begin{equation}
G_{k,q}(x,y) = \frac{1}{\mathcal{N}}\mathcal{S}\sum_{\alpha_1,\alpha_2=\pm 1}e^{i\alpha_1kx+i\alpha_2qy}\left[A(\alpha_1k,\alpha_2q)\theta(x-y)+B(\alpha_2q,\alpha_1k)\theta(y-x)\right],
\end{equation}
where $\mathcal{N}$ is the normalization factor such that $1=\sum_{x,y} |G_{k,q}(x,y)|^2$ and $\mathcal{S}$ is the symmetrizer. Note if the transformation in Eq.\eqref{Utrans} is not performed, then there is a factor $(-1)^{y}$ multiplying the wavefunction shown in the main text.
The amplitudes are related as follows:
\begin{equation}
\begin{split}
A(-k,q) &= K(k)A(k,q),\\
A(k,-q) &= K(q)A(k,q),\\
A(k,q) &= S(k,q)B(k,q),
\end{split}
\end{equation}
where $K(k)$ is given by the boundary condition
$EF(0,x)= JF(1,x) + 2i\gamma F(0,x)$ which gives
\begin{equation}
K(k) = -\frac{e^{ik}+e^{-ik}-2i\gamma-J^2e^{ik}}{e^{ik}+e^{-ik}-2i\gamma-J^2e^{-ik}},
\end{equation}
and the bulk interaction scattering matrix is that of Hubbard models given by
\begin{equation}
\label{Sint}
S(k,q) = \frac{\sin k - \sin q + 2\gamma}{\sin k - \sin q -2\gamma}.
\end{equation}
If we impose the boundary condition on a finite chain, then $k$ and $q$ are eigenstates if they satisfy the Bethe Ansatz equations
\begin{align}
e^{2ik(L+1)}&= K(k)S(k,q)S(-k,q)\\
e^{2iq(L+1)}&= K(q)S(q,k)S(-q,k),
\end{align}
which, when written explicitly, takes the form
\begin{equation}
\begin{split}
\label{SBAE}
-\frac{2\cos k-2i\gamma-J^2e^{ik}}{2\cos k-2i\gamma-J^2e^{-ik}}\left(\frac{\sin k - \sin q + 2\gamma}{\sin k - \sin q -2\gamma}\right)\left(\frac{\sin k + \sin q - 2\gamma}{\sin k + \sin q +2\gamma} \right)&= e^{i2(L+1)k}\\
-\frac{2\cos q-2i\gamma-J^2e^{iq}}{2\cos q-2i\gamma-J^2e^{-iq}}\left(\frac{\sin k - \sin q - 2\gamma}{\sin k - \sin q +2\gamma}\right)\left(\frac{\sin k + \sin q - 2\gamma}{\sin k + \sin q +2\gamma} \right)&= e^{i2(L+1)q}.
\end{split}
\end{equation}
The solution of $k$ and $q$ can be written as $k=k^R+ik^I$ and $q=q^R+iq^I$.

\subsection{1. The free band (Roots with vanishing Imaginary part)}

By taking the norm of Eq.~\eqref{SBAE}, it is easy to see that
\begin{equation}
k_I= \frac{1}{L+1} \ln | K(k)S(k,q)S(-k,q) |
\end{equation}
and the same for $k\to q$. One can see that as long as the scattering matrix is not singular, the imaginary part of the Bethe roots vanishes in the thermodynamic limit, and results in a real band of quasimomenta. In the thermodynamic limit, the imaginary part of the roots is given by replacing $k = \frac{\pi n_1}{L+1}+ik^I$ and $q = \frac{\pi n_2}{L+1}+iq^I$.

\subsection{Roots with Finite Imaginary Part}
As mentioned earlier, when the scattering matrix is singular, the Bethe roots have a non-vanishing imaginary part in the thermodynamic limit. Now, let us consider different situations with different parts of the S-matrix being singular.
\begin{figure}
\centering
\includegraphics[width=0.8\linewidth]{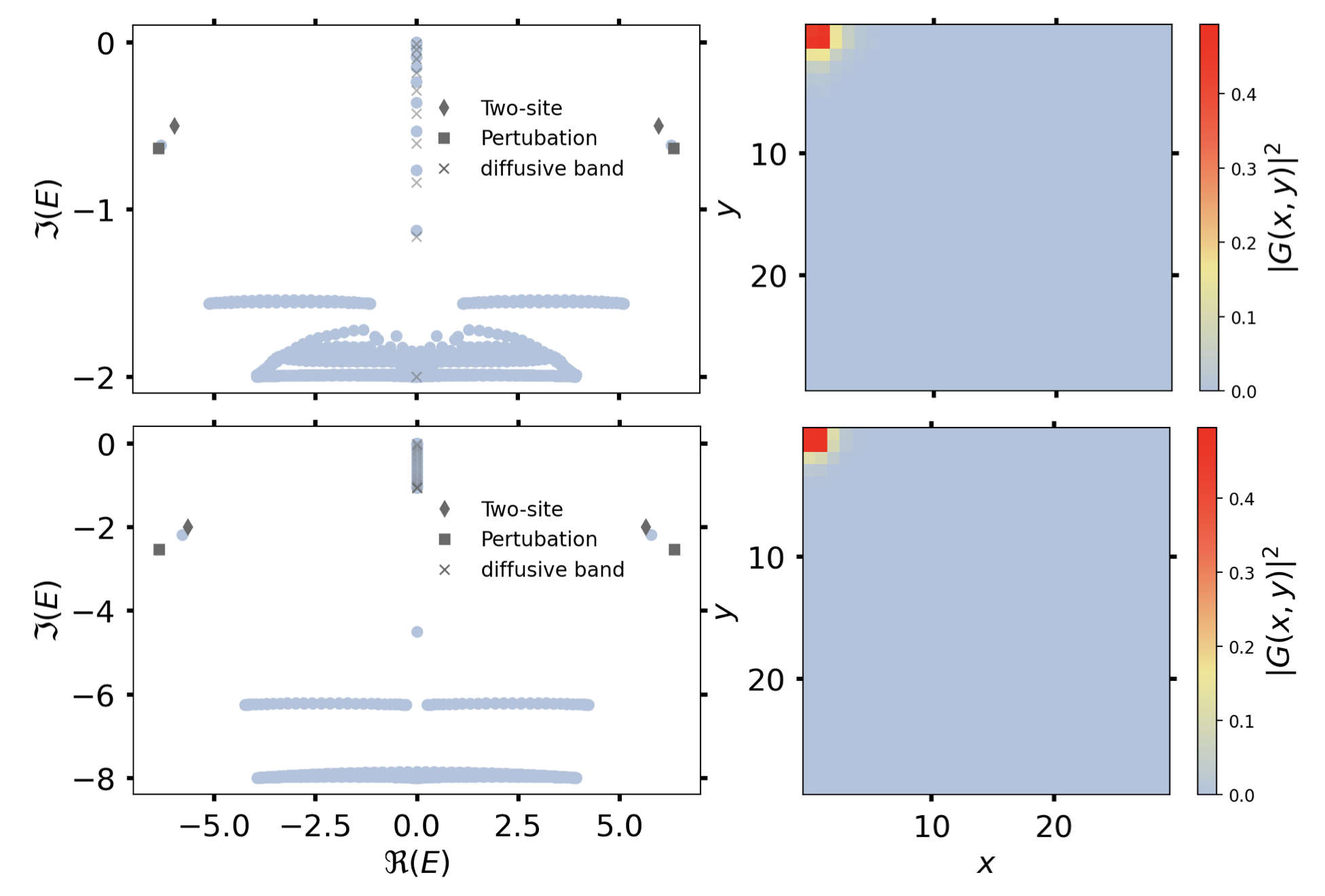}
\caption{The upper two panels show the Liouvillian spectrum (top-left) and amplitude for the wavefunction in the double-bound mode (top-right) for $J=3$, $\gamma=0.5$ and $L=30$. The perturbative calculation matches the exact diagonalization result better since $\gamma$ is small. The lower two panels show the Liouvillian spectrum (bottom-left) and amplitude for the wavefunction in the double-bound mode (bottom-right) for $J=3$, $\gamma=2$ and $L=30$. The two-site Hamiltonian calculation matches the exact diagonalization result better since $\gamma$ is big. Also, wavefunctions for both cases show that the particles are localized at the boundary more strongly and when $\gamma$ is bigger.}
\label{SBBM}
\end{figure}
\subsection{2. Boundary Mode Band}
In the case where the boundary $S-$matrix $K$ is singular, the bound mode exists, and the singular term can be written as
\begin{equation}
2\cos k-2i\gamma-J^2e^{ik}=0,
\end{equation}
which leads to the constraint $\Im k > 0$. Changing the variable to $e^{ik} = x$, the equation becomes
\begin{equation}
(1-J^2)x^2-2i\gamma x +1= 0,
\end{equation}
which has a solution
\begin{equation}
e^{ik} = \frac{\pm\sqrt{J^2-1-\gamma^2}-i\gamma }{J^2-1}:=e^{-\phi},
\end{equation}
under the assumption $|e^{ik}|>1$.
Plugging $k$ into the other equation gives the solution of $q$ as
\begin{equation}
-\frac{2\cos q-2i\gamma-J^2e^{iq}}{2\cos q-2i\gamma-J^2e^{-iq}}\left(\frac{\sin k - \sin q - 2\gamma}{\sin k - \sin q +2\gamma}\right)\left(\frac{\sin k + \sin q - 2\gamma}{\sin k + \sin q +2\gamma} \right)= e^{i2(L+1)q}.
\end{equation}
In the thermodynamic limit, $q = \frac{n_2\pi}{L+1}$ and $q = i\phi$.

\subsection{3. Diffusive Band}
In the $L\to\infty$ limit, without lack of generality, we choose $\Im k_1<0$. Then, the bulk scattering matrix being singular indicates the solutions 
satisfy the string hypothesis in the way $k_2 = k_1^*+\pi$. These solutions form a band of solutions whose corresponding eigenvalues have a vanishing real part. Now we define $k = \Re k+i \Im k$, then the S-matrix being singular also gives the relationship between $\Re k$ and $\Im k$ as 
\begin{equation}
    \Re k = \pi - \arcsin\frac{\gamma}{\cosh \Im k}.
\end{equation}
As a result, the eigenenergy is $E =2i\tanh \Im k -4i\gamma$, and the Bethe Ansatz equation can be written as the form of a Bethe-Gaudin-Takahashi (BGT) equation~\cite{alba2023free} as following
\begin{equation}
\label{BGT}
   \left(\frac{2\cos \Re k\cosh \Im k -i2\sin \Re k \sinh \Im k -2i\gamma -J^2e^{i(\Re k-\Im k)}}{2\cos \Re k\cosh \Im k -i2\sin \Re k \sinh \Im k -2i\gamma -J^2 e^{-i(\Re k-\Im k)}}\right)^2 \left(\frac{2i\sinh \Im k\cos \Re k +2\gamma }{2i\sinh \Im k\cos \Re k -2\gamma}\right)^2= e^{-i4(L+1)\Re k}.
\end{equation}
To solve this equation, it is easier to change the variable $\Im k = \mathrm{arccosh} \frac{\gamma}{z}$ and take the logarithm on both sides of the equation which gives
\begin{equation}
    2\arcsin(z_n)+\mathcal{O}(L^{-1})=-\pi \frac{n}{L}
\end{equation}
with $n=0,1,2,...,L-2$. This equation which look extremely simple gives the spectrum $ E_n = 4i\sqrt{\gamma^2-\sin\left(\frac{n\pi}{2L}\right)^2}-4i\gamma +\mathcal{O}(L^3)$, which match the numerical solution extremely well as shown in Fig.~\ref{SBBM}.
This figure shows that the analytical calculation fits the exact diagonalization reasonably well when $\gamma>1$, which the system is experiencing a true diffusive bulk. However, when $\gamma<1$, the diffusive band mixes with the free band and results in a worse but reasonable prediction because we are not taking care of the contribution from the impurity.

Noticing $\Im k<0$ gives the constraint $\Im E > -4\gamma$, and all solutions beyond this limit are not physical and should be dropped. Since $\cosh \Im k >1$, the largest imaginary energy is when $\Im k=0$, which gives $\Re k = \pi-\arcsin \gamma$ and thus resulting the energy $E_{max} = 4i\sqrt{\gamma^2-1}-4i\gamma$. This is the lower bound of the dissipation band, and when $|E_{max}|>4\gamma$, which can only happen when $\gamma<1$, the dissipation band mixes with the free band. 

\subsection{4. Double-Boundary Modes}
As discussed in the main text, there is another solution when both the boundary S-matrix and the bulk S-matrix are singular. Such a mode physically describes two particles (in the imaginary Hubbard model) that are both exponentially localized at the end of the chain, and they experience a strong imaginary Hubbard interaction. The existence of these two states is discussed in the main text with a phase diagram in Fig.\ref{Eigenvalues}(d).  However, we are not able to separate these two singularities, so we treat them in two limits. When $\gamma<1$, the double-bound mode is perturbatively connected to the case when both particles form a noiseless bound mode, which means that
\begin{equation}
    G_{k_0,k_0}(x,y) = \mathcal{N}\frac{1}{J_x}e^{-ik_0x}\frac{1}{J_y}e^{-ik_0y}
\end{equation}
with $k_0 = \frac{i}{2}\log(J^2-1)+\pi$ or $k_0 = \frac{i}{2}\log(J^2-1)$ and $\mathcal{N} = \frac{1}{J^2}+\frac{e^{-2ik_0}}{1-e^{-2ik_0}}$ being the normalization factor. Then the eigenvalue of this double-bound mode reads
\begin{equation}
    E = 4\cos k_0 -4i\gamma + 4i\gamma \left[\sum_{x=1}^L G_{k_0,k_0}(x,x)^2 + G_{k_0,k_0}(0,0)\right] +\mathcal{O}(\gamma^2),
\end{equation}
and when $\gamma>1$, the bound mode is more localized on the boundary (decays faster), as one can see from the single boundary bound mode. Then, we can treat it by solving a two-site Hamiltonian ($L=2$) as 
\begin{equation}
    h = \begin{pmatrix}
         0 & -J &-J & 0\\
         -J& -2i\gamma&  0&-J\\
         -J&  0 &-2i\gamma& -J\\
         0& -J& -J&  0
    \end{pmatrix},
\end{equation}
which has two boundary bound mode with energy $E_\pm = -i\gamma\pm\sqrt{4J^2-\gamma^2}$. It is a double-bound mode only when $\Re E\neq 0$, which gives the boundary between the Kondo phase and the bound mode phase at the large limit $\gamma$ as $J=\gamma/2$.
\section{N-particle Bethe Ansatz}\label{NparticleBA}
As mentioned earlier, the XX chain with boundary impurity maps to the open boundary condition Hubbard model with boundary impurity given by
\begin{equation}
    H=\sum_{j=1}^{L-1}\sum_{\sigma=\uparrow,\downarrow}(c^\dagger_{j,\sigma}c_{j+1,\sigma}+c^\dagger_{j+1,\sigma}c_{j,\sigma})+J(c^\dagger_{0,\sigma}c_{1,\sigma}+c^\dagger_{1,\sigma}c_{0,\sigma})+4i\gamma\sum_{j=1}^{L}n_{j,\uparrow}n_{j,\downarrow}-2i\gamma\sum_{j=1}^L\sum_{\sigma=\uparrow,\downarrow} n_{j,\sigma}.
\end{equation}

As shown in~\cite{kattel2024kondo}, the boundary scattering matrix is
\begin{equation}
    S_{b_j}(k)=\frac{-e^{2 i k_j} J^2+ e^{2 i k_j}+1}{-J^2+ \left(1+e^{2 i k_j}\right)}=e^{2ik_J}\frac{2  \cos (k_j)-i J^2 \sin (k_j)-J^2 \cos (k_j)}{2  \cos (k_j)+i J^2 \sin (k_j)-J^2 \cos (k_j)},
\end{equation}

and the Bulk S-matrix, as first calculated by Lieb and Wu for real values of U, is given by
\begin{equation}
    S_{j,l}=\frac{\sin(k_j)-\sin(k_l)-2\gamma P_{j,l}}{\sin(k_j)-\sin(k_l)-2\gamma},
\end{equation}
where $P_{j,l}$ is the permutation operator. 

Now, the usual Bethe Ansatz routine can be employed. It is very important to notice that the boundary $S-$matrix does not depend on the spin variable, but the bulk $S-$matrix depends on the spin variable (through the permutation operator). The Yang-Baxter equation is, therefore, trivially satisfied, thereby proving that the model is integrable.

Following~\cite{li2014exact,wang2015off}, we construct the N-particle Bethe Ansatz of the model using the functional Bethe Ansatz method. Let us consider the rational solution of the six-vertex model
\begin{equation}
    R_{ij}(u)=u+\eta P_{ij},
\end{equation}
where we will consider the crossing parameter $\eta=-2\gamma$.

Then, for the open boundary condition, we consider the double-row transfer matrix

\begin{align}
T_0(\lambda)&=R_{0,N}(\lambda-\sin(k_N))R_{0,N-1}(\lambda-\sin(k_{N_1}))\cdots R_{0,2}(\lambda-\sin(k_2))R_{0,1}(\lambda-\sin(k_1))\nonumber\\
\hat T_0(\lambda)&=R_{0,1}(\lambda+\sin(k_1))R_{0,2}(\lambda+\sin(k_2))\cdots R_{0,N-1}(\lambda+\sin(k_{N_1}))R_{0,N}(\lambda+\sin(k_N))\nonumber
\end{align}
Now, we define the monodromy matrix
\begin{equation}
    \Xi(\lambda)=T_0(\lambda)\hat T_0(\lambda).
\end{equation}
The trace of the monodromy matrix over the auxiliary space is defined as the double-row transfer matrix
\begin{equation}
    t(\lambda)=\operatorname{tr}_0\Xi(\lambda).
    \label{tmatrix}
\end{equation}
It is quite easy to see that the transfer matrix forms a one-parameter family of commuting operators \textit{i.e.}
\begin{equation}
    [t(\lambda),t(\rho)]=0.
\end{equation}

First, notice that using the coordinate Bethe Ansatz, one can immediately construct the eigenvalue problem for the transfer matrix
\begin{equation}
   \begin{aligned} \tau_j=
& S_{j-1, j}\left(k_{j-1}, k_j\right) \cdots S_{1, j}\left(k_1, k_j\right) {S_{b_j}}\left(k_j\right) S_{j, 1}\left(-k_j, k_1\right) S_{j, j-1}\left(-k_j, k_{j-1}\right) \nonumber\\
&S_{j, j+1}\left(-k_j, k_{j+1}\right) \cdots S_{j, N}\left(-k_j, k_N\right) S_{N, j}\left(k_N, k_j\right) \cdots S_{j+1, j}\left(k_{j+1}, k_j\right) .
\end{aligned}
\end{equation}

Upon identification of the relation between the spectral parameter valued double-row transfer matrix $t(\lambda)$ and the transfer matrix obtained from the coordinate Bethe Ansatz $\tau_j$ via
\begin{equation}
    \tau_j/S_{b_j}(k_j)=\frac{t(-\sin(k_j))}{\eta(\sin(k_j)-\eta)\prod_{l\neq j}^N(\sin(k_j)-\sin(k_l)-\eta)(\sin(k_j)+\sin(k_l)-\eta)}
    \label{cBAE-derv}
\end{equation}
Once again, the boundary $S-$matrix $S_{b_j}$ is independent of a spin variable; it is just a phase that trivially commutes with bulk $S-$matrices. 

The eigenvalue of the transfer matrix satisfies the $T-Q$ relation
\begin{align}
    \Lambda(\lambda)&=\prod_{j=1}^N\frac{2\lambda+2\eta}{2\lambda+\eta}\left(\lambda-\sin k_j+\eta\right)\left(\lambda+\sin k_j+\eta\right) \frac{Q(\lambda-\eta)}{Q(\lambda)}\nonumber\\
    &\quad+\prod_{j=1}^N\frac{2\lambda}{2\lambda+\eta}\left(\lambda-\sin k_j\right) \left(\lambda+\sin k_j\right) \frac{Q(\lambda+\eta)}{Q(\lambda)},
\end{align}
where the $Q$ function is
\begin{equation}
    Q(\lambda)=\prod_{l=1}^M (\lambda-\lambda_l)(\lambda+\lambda_l+\eta).
\end{equation}

Imposing the regularity condition of the $T-Q$ relation immediately gives the Bethe equation
\begin{equation}
    \frac{\lambda_j+\eta}{\lambda_j}\prod_{j=1}^N\frac{\lambda_j-\sin(k_j)+\eta}{\lambda_j-\sin(k_j)}\frac{\lambda_j+\sin(k_j)+\eta}{\lambda_j+\sin(k_j)}=-\prod_{l=1}^M \frac{\lambda_j-\lambda_l+\eta}{\lambda_j-\lambda_l-\eta}\frac{\lambda_j+\lambda_l+2\eta}{\lambda_j+\lambda_l}.
\end{equation}

To make the equations more symmetric, we change the variable $\lambda\to \lambda-\eta/2$, such that the Bethe equations becomes
\begin{equation}
   \frac{\lambda_j+\frac{\eta}{2}}{ \lambda_j-\frac{\eta}{2}}\prod_{j=1}^N\frac{\lambda_j-\sin(k_j)+\frac{\eta}{2}}{\lambda_j-\sin(k_j)-\frac{\eta}{2}}\frac{\lambda_j+\sin(k_j)+\frac{\eta}{2}}{\lambda_j+\sin(k_j)-\frac{\eta}{2}}=-\prod_{l=1}^M \frac{\lambda_j-\lambda_l+\eta}{\lambda_j-\lambda_l-\eta}\frac{\lambda_j+\lambda_l+\eta}{\lambda_j+\lambda_l-\eta}.
\end{equation}

From Eq.~\eqref{cBAE-derv}, upon using the shift $\lambda\to \lambda-\eta/2$ we obtain
\begin{equation}
    e^{-2ik_jL}/S_{b_j}(k_j)=\prod_{l=1}^M \frac{\sin k_j-\lambda_l+\frac{\eta}{2}}{\sin k_j-\lambda_l-\frac{\eta}{2}} \frac{\sin k_j+\lambda_l+\frac{\eta}{2}}{\sin k_j+\lambda_l-\frac{\eta}{2}}.
\end{equation}
Or,
\begin{equation}
    e^{-2ik_J(L+1)}\frac{2  \cos (k_j)+i J^2 \sin (k_j)-J^2 \cos (k_j)}{2  \cos (k_j)-i J^2 \sin (k_j)-J^2 \cos (k_j)}=\prod_{l=1}^M \frac{\sin k_j-\lambda_l+\frac{\eta}{2}}{\sin k_j-\lambda_l-\frac{\eta}{2}} \frac{\sin k_j+\lambda_l+\frac{\eta}{2}}{\sin k_j+\lambda_l-\frac{\eta}{2}}.
\end{equation}
Putting $\eta=-2\gamma$, we obtain
\begin{equation}
\label{manyparticleBAE1}
    e^{2ik_j(L+1)}\frac{2  \cos (k_j)-i J^2 \sin (k_j)-J^2 \cos (k_j)}{2  \cos (k_j)+i J^2 \sin (k_j)-J^2 \cos (k_j)}=\prod_{l=1}^M \frac{\sin k_j-\lambda_l+\gamma}{\sin k_j-\lambda_l-\gamma} \frac{\sin k_j+\lambda_l+\gamma}{\sin k_j+\lambda_l-\gamma},
\end{equation}
and the rapidities satisfy

\begin{equation}
\label{manyparticleBAE2}
    \frac{\lambda_j+\gamma}{ \lambda_j-\gamma}\prod_{j=1}^N\frac{\lambda_j-\sin(k_j)+\gamma}{\lambda_j-\sin(k_j)-\gamma}\frac{\lambda_j+\sin(k_j)+\gamma}{\lambda_j+\sin(k_j)-\gamma}=-\prod_{l=1}^M \frac{\lambda_j-\lambda_l+2\gamma}{\lambda_j-\lambda_l-2\gamma}\frac{\lambda_j+\lambda_l+2\gamma}{\lambda_j+\lambda_l-2\gamma}.
\end{equation}

If one is interested in the many-point correlation function, one needs to solve the many-particle BAE given by Eq.~\eqref{manyparticleBAE1} and Eq.~\eqref{manyparticleBAE2}.  In the main text, we only considered the two-point correlation function, and we leave the general problem of the 2n-point correlation function for future work. The Bethe equations for the real Hubbard interaction were first obtained in ~\cite{karnaukhov2005exact}.
\begin{figure}
    \centering
    \includegraphics[width=\linewidth]{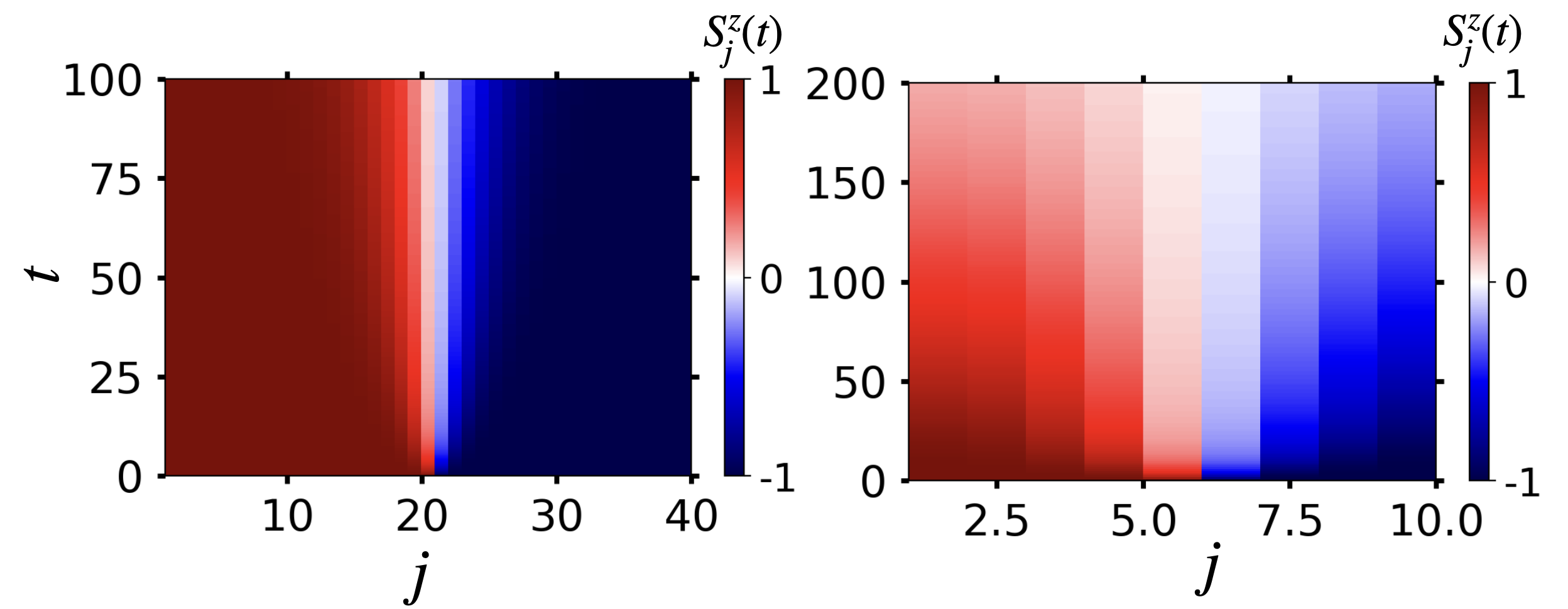}
    \caption{The magnetization $S^z_j(t)$ for time $t$ and site $j$ in the XXX case with system size $L=10$ (right panel) and XX case with system size $L=40$ (left panel) for dephasing rate $\gamma = 3$.}
    \label{S(x,t)Supp}
\end{figure}
\section{Numerical details}\label{nummmdets}
Here we will show the detailed calculation to get the diffusion constant $D$ numerically for the Heisenberg (XXX) and XX case. We start from a Doman wall configuration initially with $\rho_D = \sum_{z\in\mathcal{S}_D} \sigma^+_z\ket{0}\bra{0}\sigma^+_z$ and $\mathcal{S}_D = 0,1,2,...,L/2$. Then we evolve the state $\rho(t)$ via the Lindbladian master equation~\eqref{LeqS} from the initial state $\rho(0)=\rho_D$. Then, we calculated the magnetization profile $S^z_j(t)$ for each time $t$ and site $j$. The results for $\gamma=3$ in the XX case and XXX case are shown in Fig.~\ref{S(x,t)Supp}. For the XXX case, the domain wall hit the boundary when $t\sim 50$ and resulting in a finite size effect.

\begin{figure}[H]
    \centering
    \includegraphics[width=\linewidth]{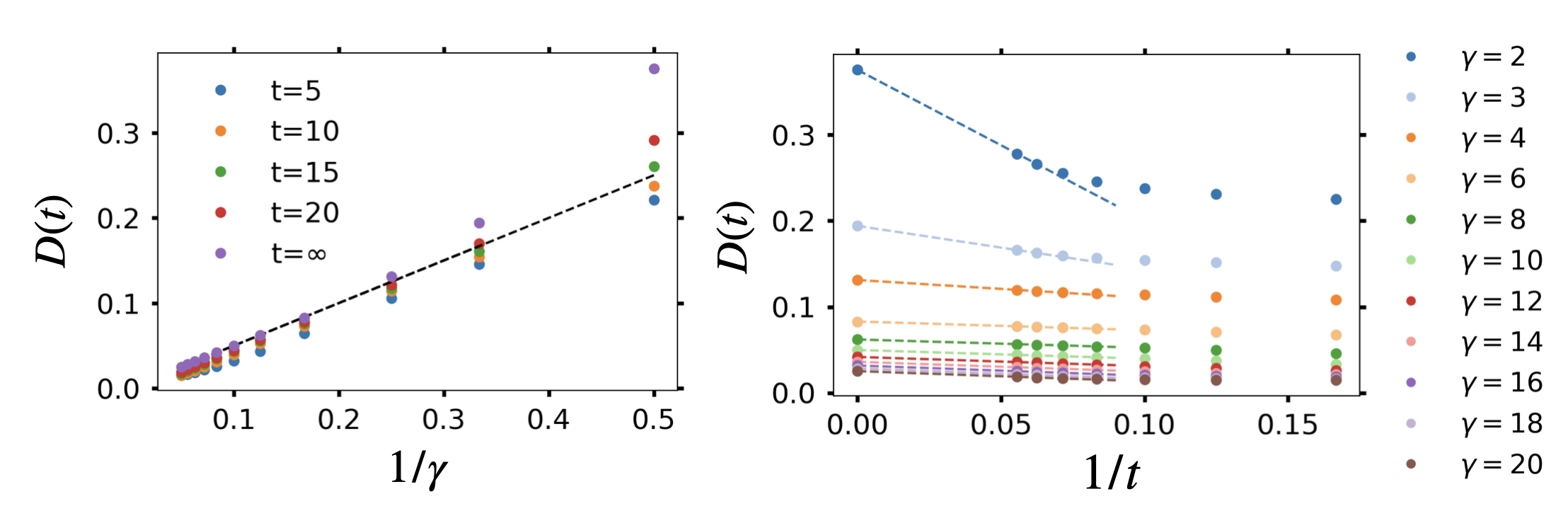}
    \caption{Scaling of the diffusion constant $D$ with time $t$ and $\gamma$ for the XXX circuit. The left panel shows $D$ as a fit parameter obtained by fitting the spin profile to the $S^z(\xi) = \mathrm{Erf}(\xi/\sqrt{D(t)})$. The right panel shows $D(t)$ as a function for different times, and we scale $D(\infty)$ by taking $1/t\to 0$. The dashed line is the fitting for $1/t\to 0$ and different colors for different $\gamma$.}
    \label{Dscale}
\end{figure}

If the bulk is diffusive, then in the limit lattice spacing $a\to0$ and system size $L\to\infty$ $S^z_j(t)\to S^z(x,t)$ should satisfy the diffusion equation $(\partial_t-D\partial_x^2)S^z(x,t)=0$. The initial condition is a domain wall meaning $S^z(x,0)=\mathrm{Sign}(L/2-x)$ and the solution reads $S^z(x,t) = -\mathrm{Erf}(\frac{x}{\sqrt{Dt}})$. It is obvious that for different times and spaces, the function collapses on the curve with a normalized variable $\xi=x/\sqrt{t}$ in the way that $S^z(\xi)=-\mathrm{Erf}(\frac{\xi}{\sqrt{D}})$. The system is not perfectly diffusive for early-time dynamics, and due to the finite-size effect, we cannot evolve the system for an extremely long time. Therefore, we
will fit the magnetization $S^z(\xi)$ to $-\mathrm{Erf}(\xi/\sqrt{D(t)})$ for a finite time $t$ to get the diffusion constant $D(t)$. To get the long-time behavior, we expand $D(t)=D(\infty)+\frac{D_1}{t}+\mathcal{O}(\frac{1}{t^2})$ and scale the diffusion constant when $t\to\infty$. The fitting and scaling are shown in Fig.\ref{Dscale}. We found for both the XX and XXX cases, $D_0$, which is the diffusion constant for long-time dynamics, is $D_0=\frac{1}{2\gamma}$ for big gamma. As for small $\gamma$, due to the finite size effect, we are not able to conclude for the XXX case. For the XX case, where we can go for larger system sizes, we can show $D_0=\frac{1}{2\gamma}$. 

\end{appendix}
\end{document}